\documentclass{aa}
\usepackage{tabularx}
\usepackage{amssymb}
\usepackage{amsmath}
\usepackage{txfonts}
\usepackage{balance}
\usepackage{natbib}
\bibpunct{(}{)}{;}{a}{}{,} 
\usepackage{xspace}
\usepackage[normalem]{ulem} 
\usepackage[svgnames,rgb,table]{xcolor} 
\usepackage{multirow}
\usepackage{rotating}
\usepackage{bbm}
\usepackage{cancel}
\usepackage{graphicx}
\usepackage{mathabx}
\usepackage{caption}
\usepackage{ifthen}
\usepackage{booktabs}
\usepackage{tabularx}
\usepackage{float}

\newcommand{\del}{\partial}

\newcommand{\St}{\text{St}\xspace}
\newcommand{\rhos}{\ensuremath{\rho_\text{s}}\xspace}
\newcommand{\rhodust}{\ensuremath{\rho_\mathrm{d}}\xspace}
\newcommand{\rhogas}{\ensuremath{\rho_\mathrm{g}}\xspace}

\newcommand{\uf}{\ensuremath{u_\text{f}}\xspace}
\newcommand{\csound}{\ensuremath{c_\mathrm{s}}\xspace}

\newcommand{\Hg}{\ensuremath{H_\mathrm{g}}\xspace}
\newcommand{\Hd}{\ensuremath{H_\mathrm{d}}\xspace}

\newcommand{\Ok}{\ensuremath{\Omega_\mathrm{k}}\xspace}

\newcommand{\Siggas}{\ensuremath{\Sigma_\mathrm{g}}\xspace}
\newcommand{\Sigdust}{\ensuremath{\Sigma_\mathrm{d}}\xspace}

\newcommand{\alphat}{\ensuremath{\alpha_\text{t}}\xspace}
\newcommand{\ugas}{\ensuremath{u_\text{gas}}\xspace}

\newcommand{\dlnPdlnR}{\ensuremath{ {\left| \frac{\mathrm{d}\ln P}{\mathrm{d}\ln r} \right| } }\xspace}
\makeatletter
\if@referee
    \newcommand{\change}[2]{\sout{#1}\xspace\textbf{{#2\xspace}}}  
    \usepackage[nolists,noheads,nomarkers]{endfloat}               
\else
    \newcommand{\change}[2]{#2\xspace}                             
\fi
\makeatother
\usepackage[%
pdfauthor={Tilman Birnstiel},
pdftitle={A simple model for the evolution of the dust population in protoplanetary disks},
pdfstartview=FitH,
linkcolor=blue,
anchorcolor=blue,
citecolor=blue,
filecolor=blue,
menucolor=blue,
urlcolor=blue,
colorlinks=true]{hyperref}
\usepackage{pdfsync}
\usepackage[draft,color=yellow,author={Til Birnstiel}]{pdfcomment}
\newcommand{\highlight}[1]{\textcolor{black}{#1}}

\begin{document}
\title{A simple model for the evolution of the dust population in protoplanetary disks}
\titlerunning{A simple model for dust evolution in protoplanetary disks}
\author{T.~Birnstiel\inst{1}\fnmsep\inst{2} \and H.~Klahr\inst{3} \and B.~Ercolano\inst{1}\fnmsep\inst{2}}
\authorrunning{T.~Birnstiel et al.}
\institute{
    University Observatory Munich, Scheinerstr. 1, D-81679 M\"unchen, Germany
    \and
    Excellence Cluster Universe, Boltzmannstr. 2, D-85748 Garching, Germany
    \and
    Max-Planck-Institut f\"ur Astronomie, K\"onigstuhl 17, D-69117 Heidelberg, Germany
}
\date{\today}

\abstract
{The global size and spatial distribution of dust is an important ingredient in the structure and evolution of protoplanetary disks and in the formation of larger bodies\highlight{, such as planetesimals}.
}
{We aim to derive simple equations that explain the global evolution of the dust surface density profile and the upper limit of the grain size distribution and which can readily be used for further modeling or for interpreting of observational data.}
{We have developed a simple model that follows the upper end of the dust size distribution and the evolution of the dust surface density profile. This model is calibrated with state-of-the-art simulations of dust evolution, which treat dust growth, fragmentation, and transport in viscously evolving gas disks.}
{We find very good agreement between the full dust-evolution code and the toy model presented in this paper.
We derive analytical profiles that describe the dust-to-gas ratios and the dust surface density profiles well in protoplanetary disks, \highlight{as well as the radial flux by solid material ``rain out'', which is crucial for triggering any gravity assisted formation of planetesimals}.We show that fragmentation is the dominating effect in the inner regions of the disk leading to a dust surface density exponent of ${-1.5}$, while the outer regions at later times can become drift-dominated, yielding a dust surface density exponent of $-0.75$. Our results show that radial drift is not efficient in fragmenting dust grains. This supports the theory that small dust grains are resupplied by fragmentation due to the turbulent state of the disk.}
{}


\maketitle

\section{Introduction}\label{sec:introduction}
The dust content is fundamental for many aspects of the evolution, structure, and observations of protoplanetary disks. Not only is this the material out of which terrestrial planets and the cores of giant planets form, but it also dominates the opacity, thereby determining the temperature structure and the observational appearance of protoplanetary disks. The grain surfaces also provide the ground for chemical surface reactions \citep[e.g.,][]{Aikawa:2006p15367} and electron capture, so that dust strongly influences the ionization state and possibly the turbulent state of the disk \citep{Sano:2000p9889,Bai:2009p15295,Turner:2010p15252}.

\begin{figure*}[tb]
  \centering
  \resizebox{!}{0.283\hsize}{\includegraphics{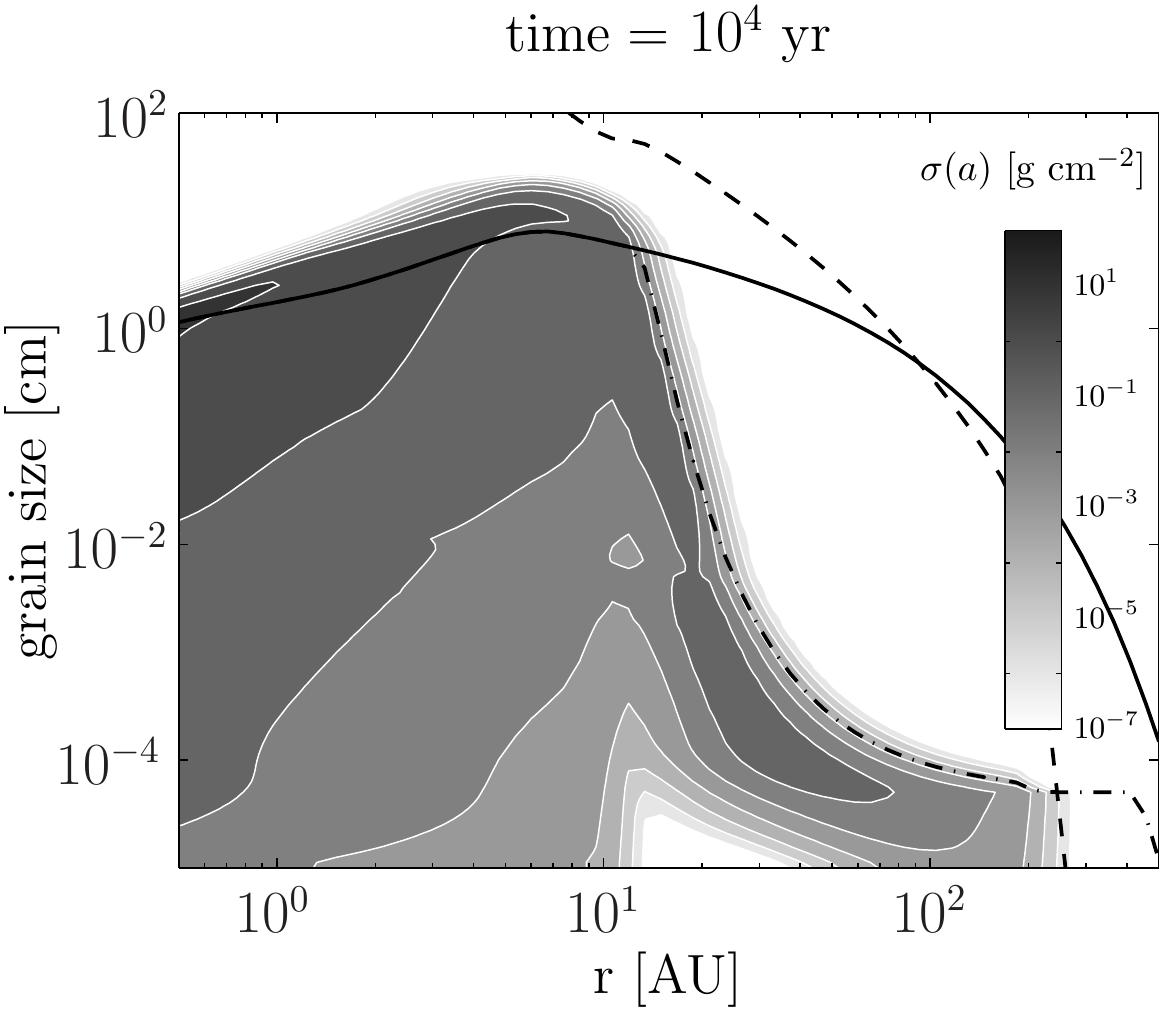}}
  \resizebox{!}{0.28\hsize}{\includegraphics{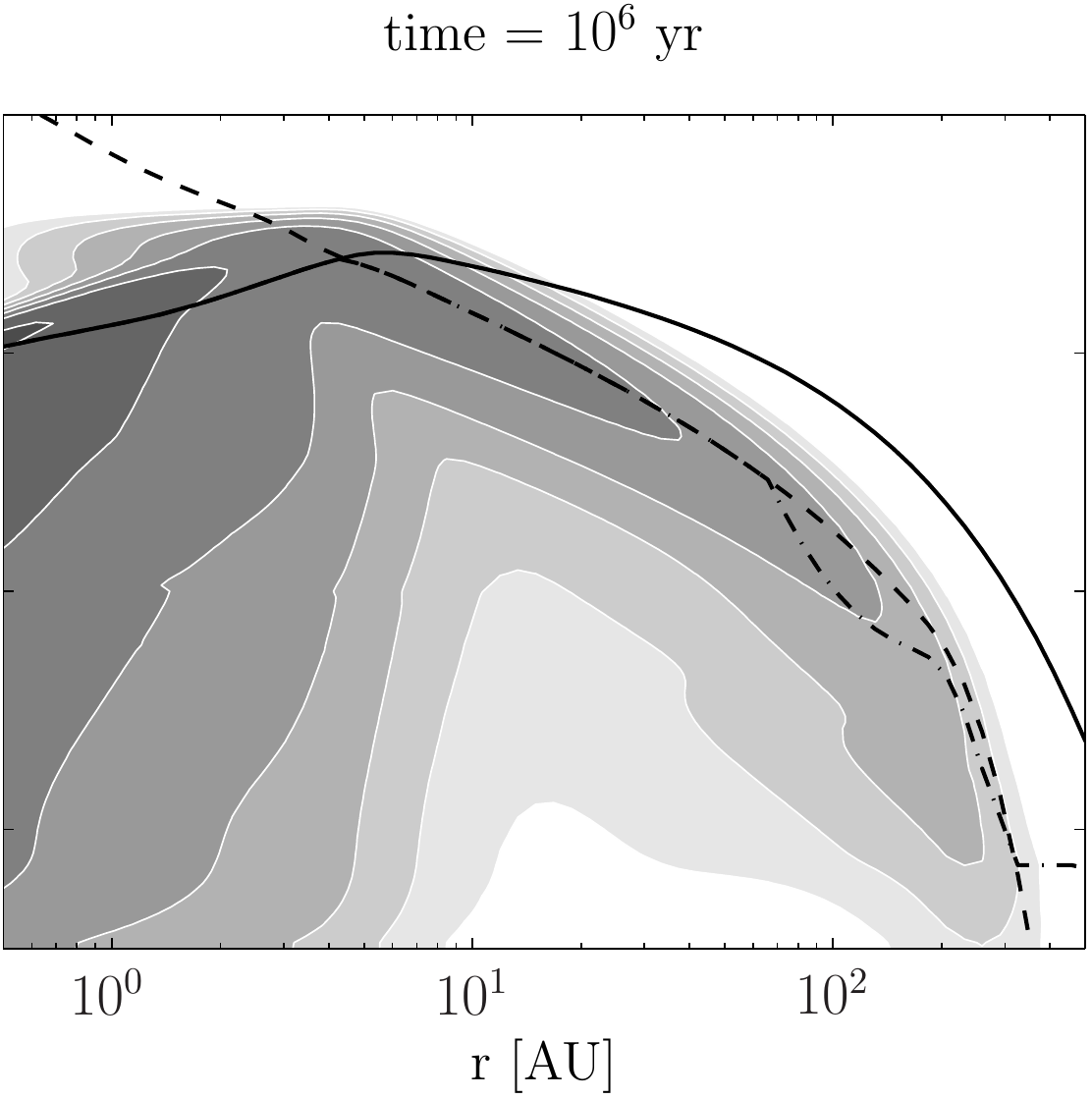}}
  \resizebox{!}{0.28\hsize}{\includegraphics{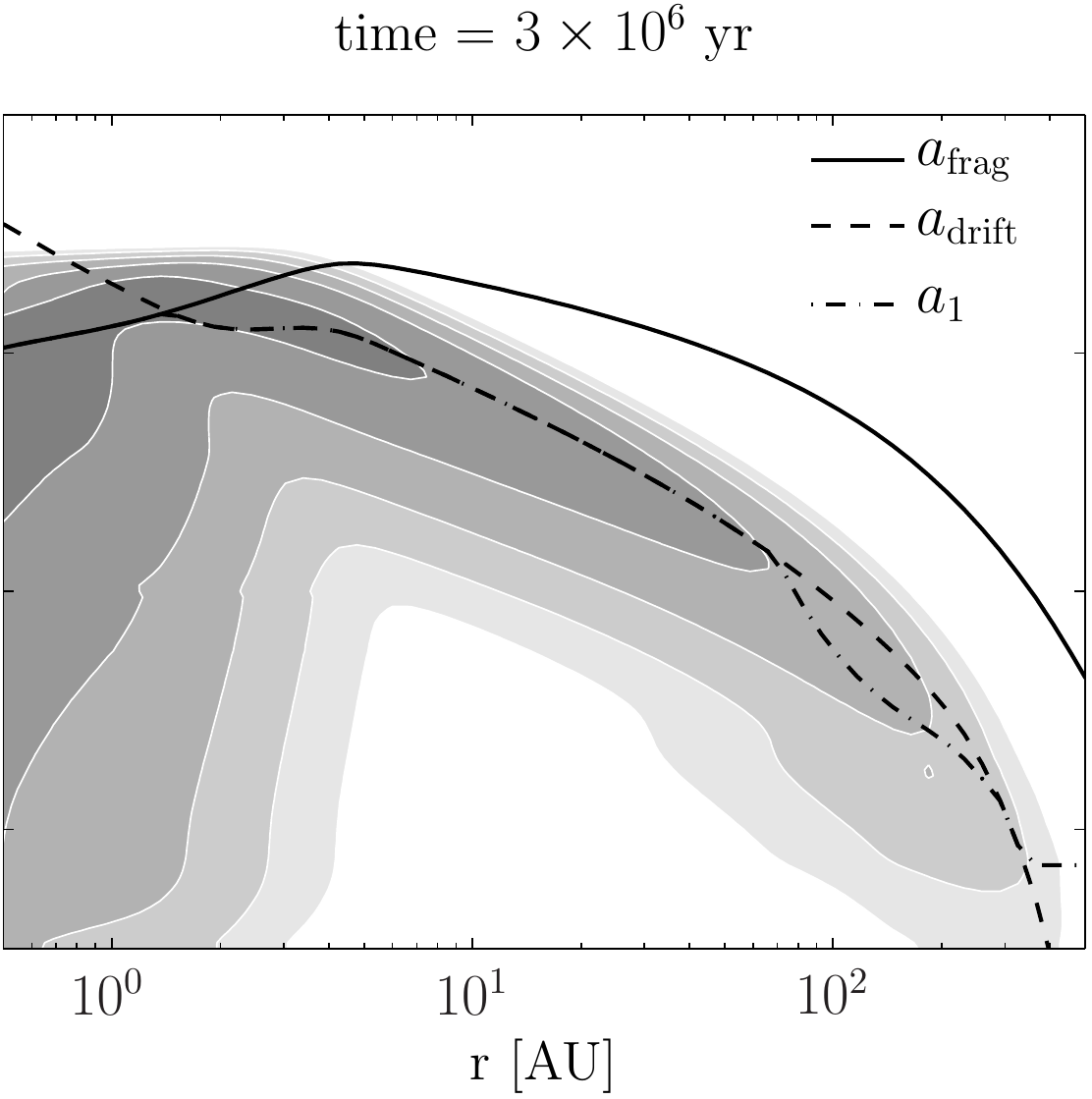}}
  \caption{Dust surface density distributions as function of radius and grain size at $10^4$, $10^6$, and $3\times 10^6$ years. In this simulation, the turbulence parameter $\alphat$ was taken to be $10^{-3}$. Overlayed are the representative sizes for a fragmentation limited size distribution (solid black lines) and for a drift limited size distribution (dashed black lines) according to Eq.~\ref{eq:a_frag} and Eq.~\ref{eq:a_drift}, respectively. The dash-dotted line denotes the time-dependent estimate of the largest grain size $a_1(t)$ as defined by Eq.~\ref{eq:a_1_of_t}.} 
  \label{fig:sim_results_A}
\end{figure*}

While there have been studies of gas line emission \citep[e.g.,][]{Dutrey:2007p15375,Panic:2008p5086}, most disk mass estimates are based on observations of dust emission, assuming a constant dust-to-gas ratio \citep[see][]{Andrews:2005p3779,Andrews:2007p3380}. However, it is well known that the dust-to-gas ratio in disks should be an evolving quantity \citep[e.g.,][]{Weidenschilling:1977p865,Nakagawa:1981p4533,Keller:2004p867,Brauer:2007p232,Ciesla:2009p10132}. The problem with our understanding of dust transport is that it depends on the structure of the disk, which is uncertain, and also on the size distribution of the grains, which is also poorly understood.

Several authors have investigated simplified models of dust transport and size evolution: for example, \citet{Garaud:2007p405}, \citet{Kornet:2001p688}, \citet{Rozyczka:2004p15357}, \highlight{\citet{Klahr:2006p7719}}, or \citet{Hughes:2010p15360}. A more complete picture is given by global simulations that self-consistently treat both the transport and collisional evolution, such as the works by \citet{Weidenschilling:1997p4593}, \citet{Brauer:2008p212}, \citet{Birnstiel:2010p9709}, or \citet{Okuzumi:2011p15378}. The last models (which we will call \emph{full models} as opposed to the \emph{simplified models}, such as the one presented in this paper), however, are often so complicated that first it is difficult to derive a physical understanding from them, and second they are computationally expensive.

We therefore put forward a model that is motivated by and based on the full simulations mentioned before \citep[in this case, ][]{Birnstiel:2010p9709}; however, it is very simplified so easier to understand, easier to implement, and computationally much less intensive. The aim of this work is to identify a few equations that govern the spatial evolution of dust as found in the full simulations. We derive upper limits for the size distribution from which we can estimate the drift velocity of the largest particles. The size distribution is then represented by just two grain sizes, one size representing the small dust that is coupled to the gas and another size representing the largest particles, which can be drifting inwards at a much higher velocity. While mathematically similar, the basic approach of this work is thus quite different from other simplified models \citep[e.g.][]{Kornet:2001p688,Garaud:2007p405}: we do not estimate the outcome of the physical processes, but instead identify the important concepts in the full models to construct a simplified model that is calibrated to reproduce the outcome of the full model.

\highlight{\citet{Klahr:2006p7719} have estimated how much solid material can be concentrated in decimeter-sized radially drifting particles in local dust traps. Once these over-densities reach Toomre unstable values and exceed the local Roche density, planetesimal formation via self gravity in the dust layer is the consequence \citep{Johansen:2006p7466}. The latter mechanism has so far only be demonstrated for ad hoc local dust to gas ratios and size distributions. What is missing is a dust evolution model, as the one presented in this work, which predicts a local dust flux and the size distribution for models of gravoturbulent planetesimal formation \citep{Klahr:2008p15769}.}

The structure of this paper is as follows: in Section~\ref{sec:background}, we introduce a few concepts and formulae that are instrumental for the rest of this work. Section~\ref{sec:sim_results} presents some fiducial simulation results of the complete model that are used in Section~\ref{sec:two_pop_model} to derive and test a simple model of dust transport. Section~\ref{sec:discussion} makes use of this model to discuss and explain some of the implication of it, such as the observational significance of the model, which profiles of the dust surface density or the dust-to-gas ratio are to be expected under certain physical conditions, as well as the resulting dust mass flux. Our findings are summarized in Section~\ref{sec:summary}.

\section{Background}\label{sec:background}

Dust particles in protoplanetary disks are subject to radial drift (which depends on the pressure structure of the disk, see \citealp{Weidenschilling:1977p865}), gas drag (i.e., particles following the gas accretion flow) and turbulent mixing. These effects depend on the size of the particles, which is also an evolving quantity because grains grow by sticking collisions. However there are several mechanisms that can limit collisional growth of dust particles: charging effects \citep{Okuzumi:2009p7473}, bouncing \citep{Guttler:2010p9745,Zsom:2010p9746}, fragmentation and radial drift \citep{Weidenschilling:1977p865,Brauer:2008p215,Birnstiel:2010p9709}.

The charging and bouncing barriers are very much dependent on the material, porosity, or fractal dimension of the grains, and are still strongly debated topics \citep[e.g.,][]{Wada:2011p15328}. In one way or another these barriers need to be overcome in order to explain the existence of larger bodies such as asteroids or planets. The fact that disks are observed to be rich in small dust particles \citep[for a recent review, see][]{Williams:2011p15776} tells us that fragmentation must be effective in wide regions of the disk, otherwise coagulation would quickly render the disk optically thin even in the infrared wavelength regime \citep[see][]{Dullemond:2004p390,Birnstiel:2009p7135}. We take this observational fact as motivation for only considering fragmentation/cratering and radial drift as growth-limiting mechanisms \citep{Dominik:2008p4626}. For describing both of these effects, it is convenient to define the Stokes number \St, which is the ratio of the stopping time of a particle $t_\text{stop}$ and the turn-over time of the largest eddy in the turbulent environment,
\begin{equation}
\St = \frac{t_\text{stop}}{t_\text{L}},
\label{eq:def_stokesnumber}
\end{equation}  
which under typical assumptions (Epstein drag law, compact spherical particles, isothermal gas density profile, and $t_\mathrm{L}= \Ok^{-1}$, see \citealp{Cuzzi:2001p2167}) simplifies for particles near the disk mid-plane to
\begin{equation}
\St = \frac{a \, \rhos}{\Siggas} \, \frac{\pi}{2}.
\end{equation}
Here $a$ is the radius and \rhos the internal density of the dust aggregate (taken to be 1.6 g cm$^{-3}$) and \Siggas the gas surface density. The Stokes number is a dimensionless quantity that describes the aerodynamic properties of a dust particle and thus its coupling to the gas flow, i.e. particles with different sizes, densities, or shapes but the same Stokes number are aerodynamically identical. 

\begin{figure*}[tb]
  \centering
  \resizebox{!}{0.283\hsize}{\includegraphics{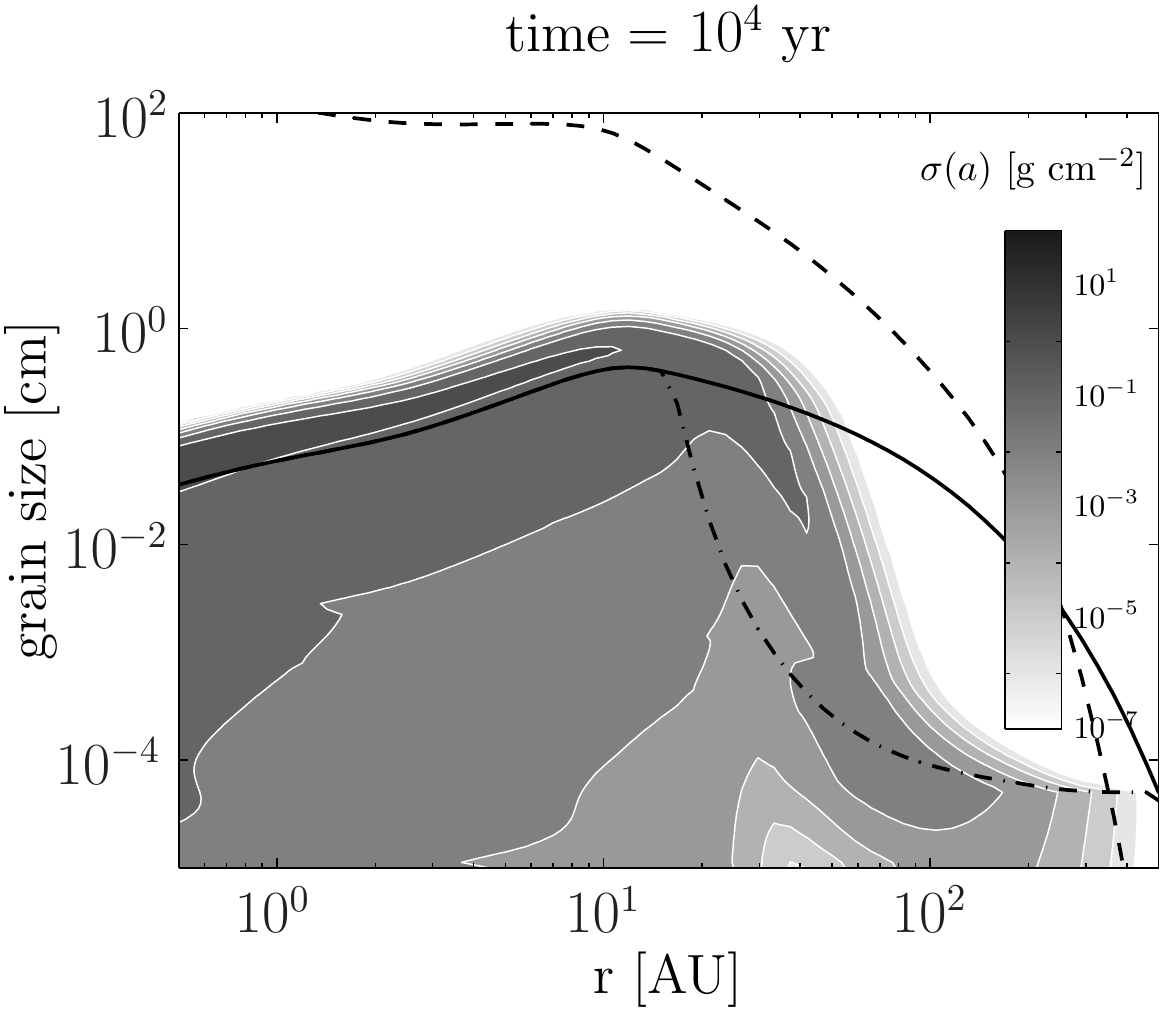}}
  \resizebox{!}{0.28\hsize}{\includegraphics{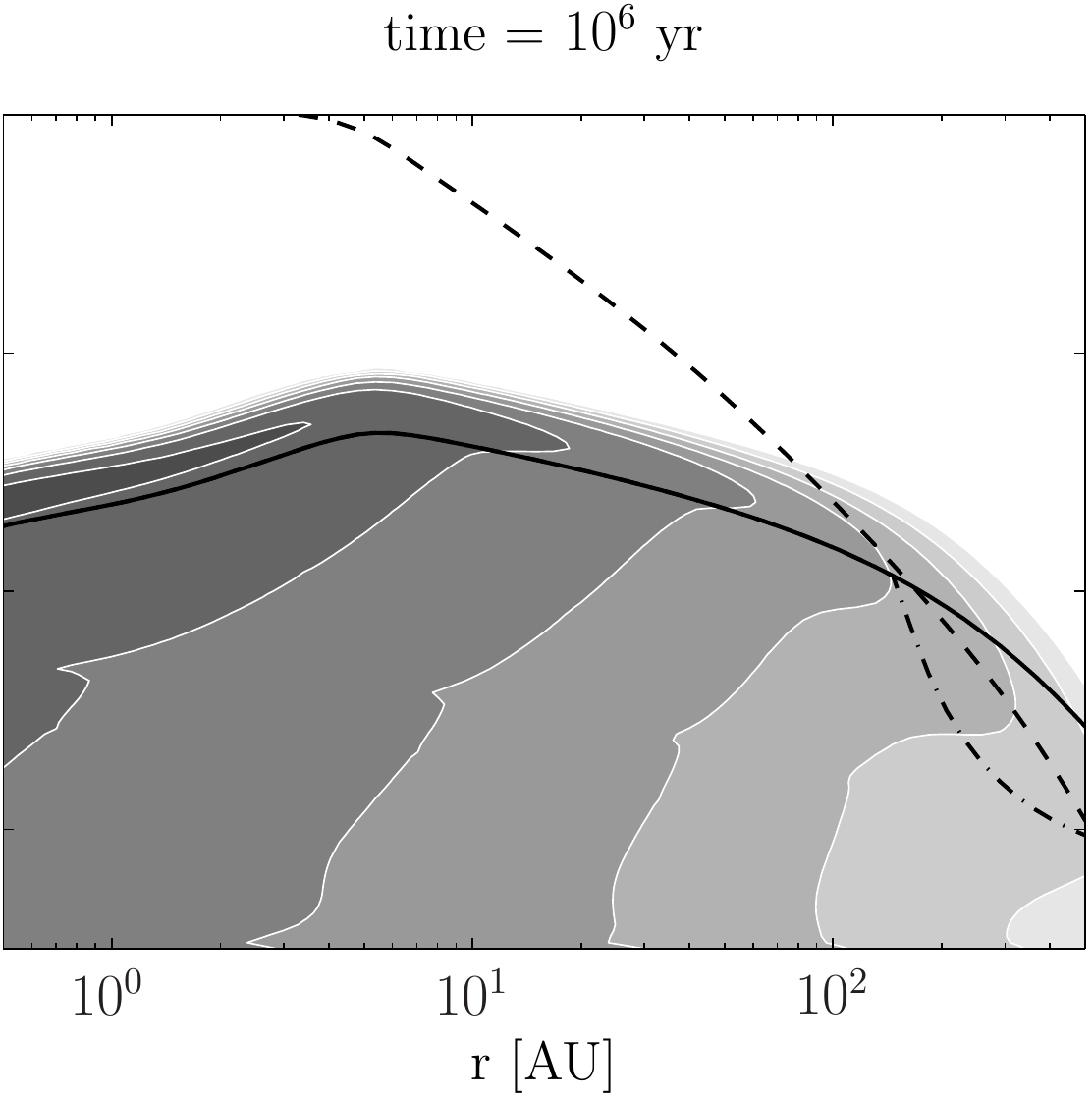}}
  \resizebox{!}{0.28\hsize}{\includegraphics{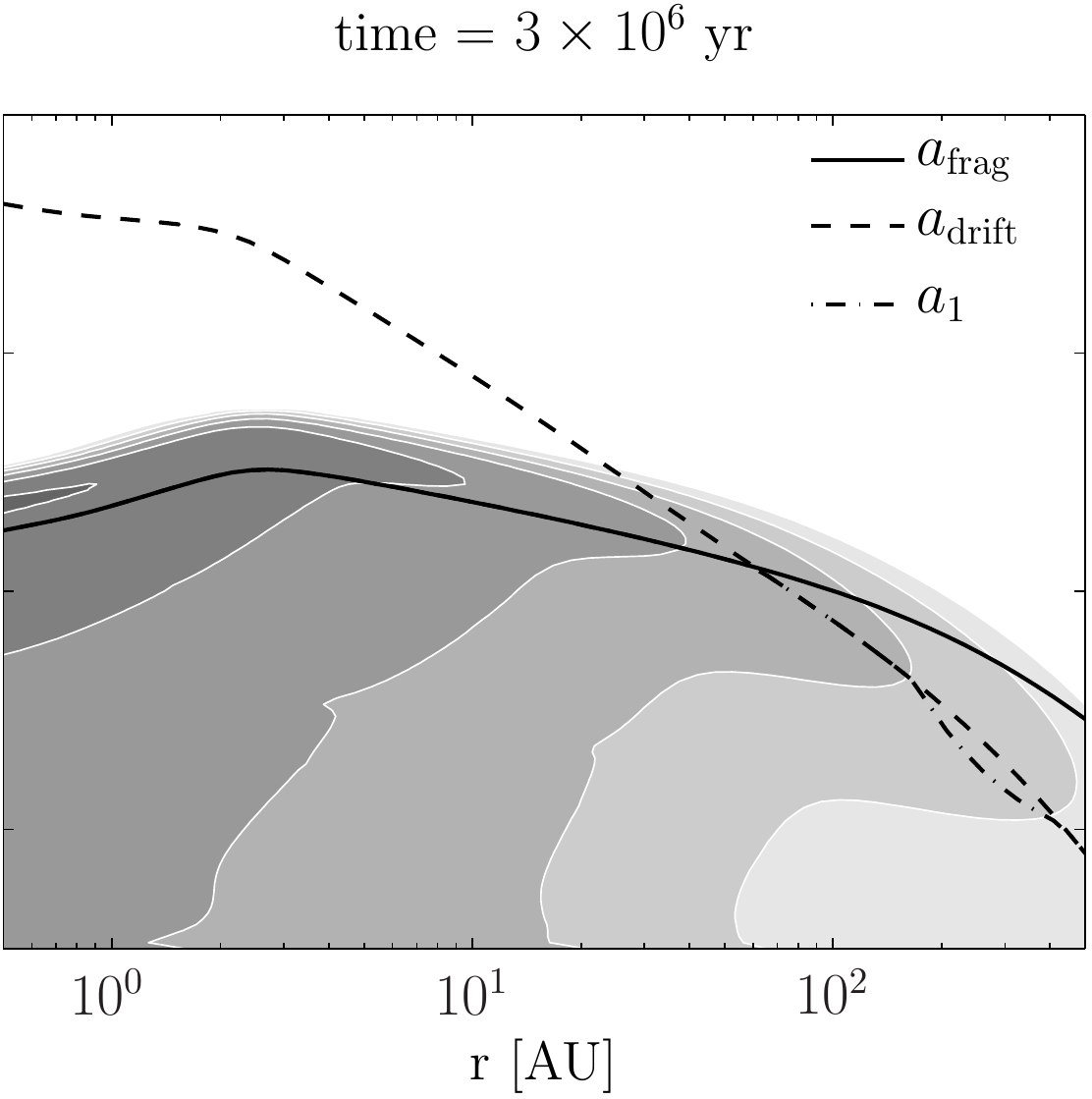}}
  \caption{As in Fig.~\ref{fig:sim_results_A} but with a higher turbulence parameter $\alphat=10^{-2}$.} 
  \label{fig:sim_results_B}
\end{figure*}

Fragmentation of dust particles stops further growth because relative velocities tend to increase with the Stokes number of the particle (for $\St<1$). They can reach values well above the fragmentation threshold velocity \uf \citep{Brauer:2008p215}.
Lab experiments measured threshold velocities of around 1~m~s$^{-1}$ for the onset of fragmentation of silicate dust grains \citep[e.g.,][]{Blum:2008p1920}. It has been suspected that icy particles fragment only at higher velocities due to the $\sim$10 times higher surface energies \citep{Wada:2008p4903,Gundlach:2011p15761}.
Recent numerical studies of \citet{Wada:2009p8776} found that the fragmentation velocity for $\lesssim$10$\mu$m sized icy aggregates could be as high as 50 m~s$^{-1}$. However, experiments with silicate dust grains find a fragmentation velocity that decreases with grain size \citep{Beitz:2011p16199}, which can at least partly be attributed to the growing importance of inhomogeneities in larger grains \citep{Geretshauser:2011p16677}.
In the following we discuss mostly results for a threshold velocity of 10~m~s$^{-1}$, since we are interested in the \textit{global} evolution of dust, i.e. the regions beyond the ice line. The Stokes number, at which the typical impact velocity of similar-sized grains equals the fragmentation threshold velocity can be approximated by \citep[see][]{Birnstiel:2009p7135}
\begin{equation}
\St_\mathrm{frag} = \frac{1}{3} \, \frac{\uf^2}{\alphat \, \csound^2},
\label{eq:St_f}
\end{equation}
where \alphat is the turbulence parameter \citep{Shakura:1973p4854} and \csound is the isothermal sound speed and we assume $\St<1$. The value of \alphat is still largely unknown, but observations of accretion signatures or disk lifetimes suggest values between $10^{-3}$ and $10^{-2}$. Theoretical models of the magnetorotational instability tend to reproduce these ranges of values \citep[e.g.][]{Johansen:2005p8425,Dzyurkevich:2010p11360}, but global simulations also suggest a radial dependence of \alphat \citep[see][]{Sano:2000p9889,Flock:2011p15542}. The simulations shown in this paper were carried out with radially constant values of \alphat, but the model introduced in this work is able to treat also radial variations of \alphat.

Apart from fragmentation, radial drift can also be a limiting factor for grain growth. Radial inward drift of dust particles occurs because the head wind induced by the sub-Keplerian gas disk removes angular momentum from the dust particles. The drift velocity \citep{Weidenschilling:1977p865}
\begin{equation}
u_\text{D} = -\frac{2 \, u_\eta}{\St + \St^{-1}},
\label{eq:u_drift}
\end{equation}
thus depends on the gas pressure gradient since $u_\eta$ is the difference between the orbital velocity of the dust grain and the gas (i.e., the head wind velocity), which is given by \citep[see][]{Weidenschilling:1977p865,Nakagawa:1986p2048}
\begin{equation}
u_\eta = - \frac{\partial P}{\partial r} \, \frac{1}{2 \rhogas \Ok}.
\label{eq:u_eta}
\end{equation}
Here $P$ denotes the gas pressure, \rhogas the mid-plane gas density, and \Ok the Keplerian angular velocity. 
\highlight{It is straight-forward to also include reduction factors of the radial drift speed in Eq.~\ref{eq:u_drift}, as found in \citet{Johansen:2006p7466}, however this is not the subject of this work.}

\section{Typical simulation results}\label{sec:sim_results}
\begin{figure}[tb]
  \centering
  \resizebox{\hsize}{!}{\includegraphics{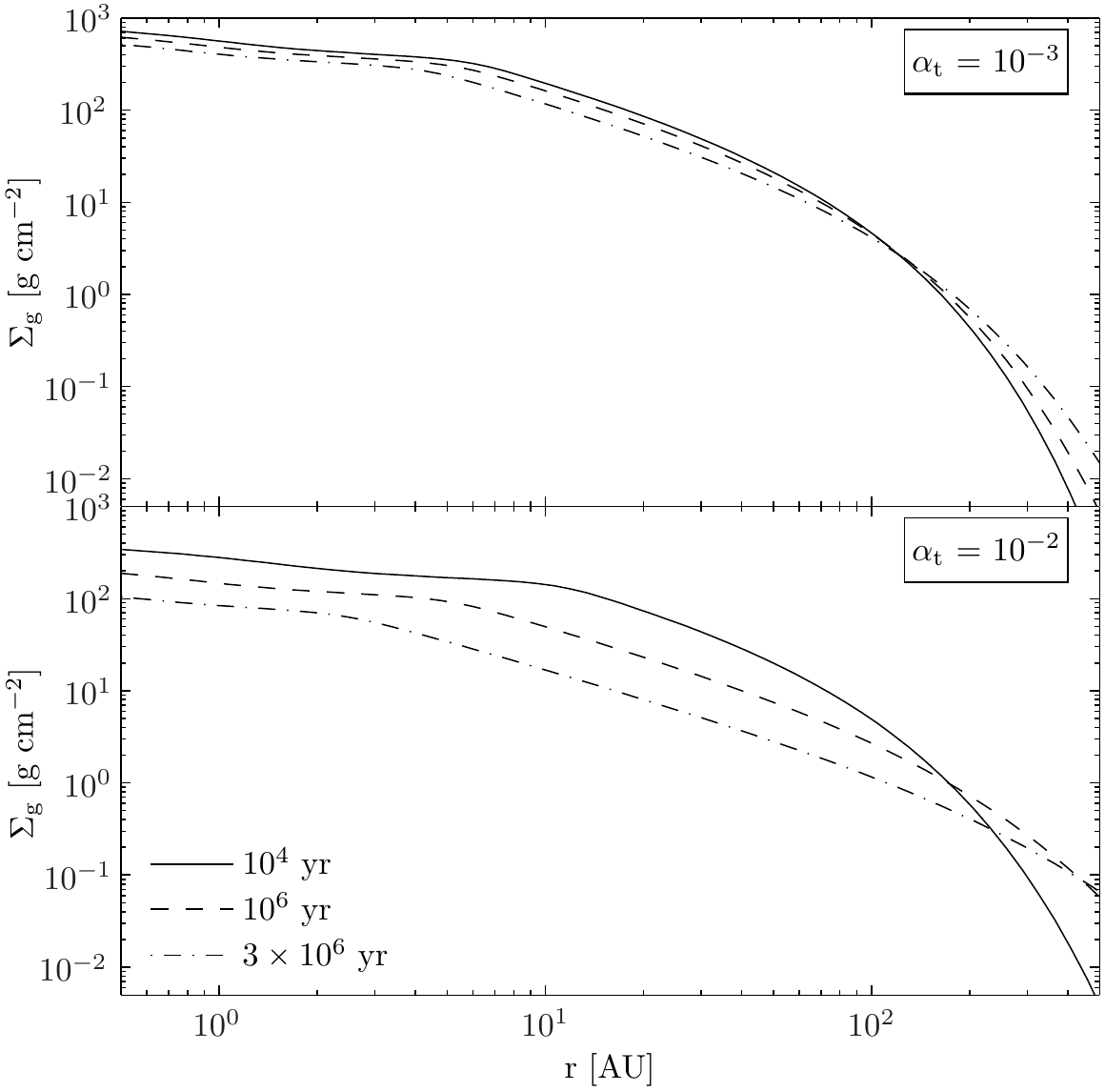}}
  \caption{Evolution of the gas surface density profiles of the low turbulent (top panel) and the high turbulent simulation (bottom panel). The solid, dashed, and dash-dotted lines correspond to simulation times of $10^4$, $10^6$, and $3\times 10^6$, respectively.}
  \label{fig:sigma_g}
\end{figure}

In this section we review some typical simulation results of the complete model that treats dust growth, fragmentation and cratering as well as radial transport due to gas drag, radial drift and turbulent mixing. These results are used in Section~\ref{sec:two_pop_model} to derive a simple model that reproduces the overall evolution of the dust surface density. The gas surface density is viscously evolving using the typical turbulent viscosity prescription \citep{Shakura:1973p4854}. The temperature is estimated taking passive irradiation and viscous heating into account. A detailed description of the model can be found in \citet{Birnstiel:2010p9709}. Figures~\ref{fig:sim_results_A} and \ref{fig:sim_results_B} show snapshots from simulations of a 0.1~$M_\odot$ disk around a solar mass star. The contours denote $\sigma(a,r)$, the dust surface density distribution per logarithmic size bin, which is defined such that 
\begin{equation}
\Sigdust(r) = \int_{a=0}^\infty \sigma(a,r) \, \text{dln} a
\end{equation}
is the total dust surface density. Throughout this paper we define $r$ as the cylindrical distance to the central star.
The initial gas surface density profile for the simulations is close to the self-similar solutions from \citet{LyndenBell:1974p1945}
\begin{equation}
\Siggas(r) \propto \left( \frac{r}{r_\mathrm{c}} \right)^{-p} \, \exp\left[-\left(\frac{r}{r_c}\right)^{2-p} \right]
\label{eq:sig_gas}
\end{equation}
for a viscosity power-law index $p$ of unity and a characteristic radius $r_\mathrm{c}$ of 60~AU (see Fig.~\ref{fig:sigma_g}). The only significant deviation is in the inner parts of the disk where the viscous heating leads to a steeper temperature profile and consequently to a flatter surface density profile. Outside of about 3~AU, there is no deviation $>30$\% between the simulations and the self-similar solution. The evolution of the gas surface density for both simulations is shown in Fig.~\ref{fig:sigma_g}.
 
Figures~\ref{fig:sim_results_A} and~\ref{fig:sim_results_B} differ only in the turbulence parameter \alphat, which is $10^{-3}$ in the former and $10^{-2}$ in the latter case.
There are a few fundamental features that are important for the further understanding of this work:
\begin{itemize}
\item 
in the left panels of Fig.~\ref{fig:sim_results_A} and \ref{fig:sim_results_B} it can be seen that grains in the inner regions quickly grow and reach a maximum size, the so-called fragmentation barrier, which is discussed in Section~\ref{sec:two_pop_model:size_limits:turb_frag}. The size limit set by fragmentation is shown as black solid lines\footnote{More precisely, the solid line is the representative size in a fragmentation limited distribution that is by definition slightly below the upper end of the size distribution, as can be seen in Fig.~\ref{fig:sim_results_A}.}.
A steady state is reached in which fragmentation and coagulation balance each other. We call these regions \emph{fragmentation limited}. As we go to larger radii in the first panel, we see that grains at about 20~AU and beyond are still at the state of initial growth, \change{}{as discussed in Section~\ref{sec:two_pop_model:particle_growth}}. In the central panel of Fig.~\ref{fig:sim_results_B} we see a similar behavior, but growth does happen on slightly shorter time scales because the growth time scale of small particles ($\St<\alphat$) is shorter due to the higher turbulent velocities. It is only for larger particles that the growth time scale does not depend on the turbulence strength \citep{Brauer:2008p215}. 
\item
In the central panels of Figs.~\ref{fig:sim_results_A} and \ref{fig:sim_results_B}, we can see that grains also in the outer regions have grown significantly. The size distribution in the inner regions are basically unchanged compared to earlier times since they are still in coagulation-fragmentation equilibrium. For reasons which are discussed later, the barriers to growth, shown as dashed and solid black lines in Figs.~\ref{fig:sim_results_A} and \ref{fig:sim_results_B} also evolve with time. Most significant is the fact that the drift limit (dashed lines, discussed in Section~\ref{sec:two_pop_model:size_limits:rad_drift}) is moving down to smaller sizes. 
\item
In the outer regions at later times, the drift barrier (dashed line) has moved below the fragmentation barrier (solid line) as can be seen in the central and right panels of Fig.~\ref{fig:sim_results_A}. This means that dust particles in those regions are no longer limited by fragmentation, but instead they are more rapidly drifting away than being replenished by growth from smaller sizes. We call those regions \emph{drift limited}.
\item
Due to the higher strength of turbulence in Fig.~\ref{fig:sim_results_B}, the fragmentation limit shifted to smaller sizes compared to Fig.~\ref{fig:sim_results_A}, which renders the disk fragmentation limited up to about 70~AU, while in Fig.~\ref{fig:sim_results_A}, only the inner regions up to about 2~AU are fragmentation-dominated.
\end{itemize}

\section{The two-population model}\label{sec:two_pop_model}
In this section, we develop an analytical understanding of dust evolution in size and spatial dimension. The simulation results of the comprehensive model, which have been introduced in the previous section are used to derive simple equations that characterize the size distribution. This enables us to develop a simple model that can reproduce the evolution of the dust surface density of the full-scale simulations well. While the following section comprehensively explains the derivation of the model, Appendix~\ref{app:algorithm} summarizes the model in a concise algorithmic form.   

\begin{figure}[tb]
  \centering
  \resizebox{0.9\hsize}{!}{\includegraphics{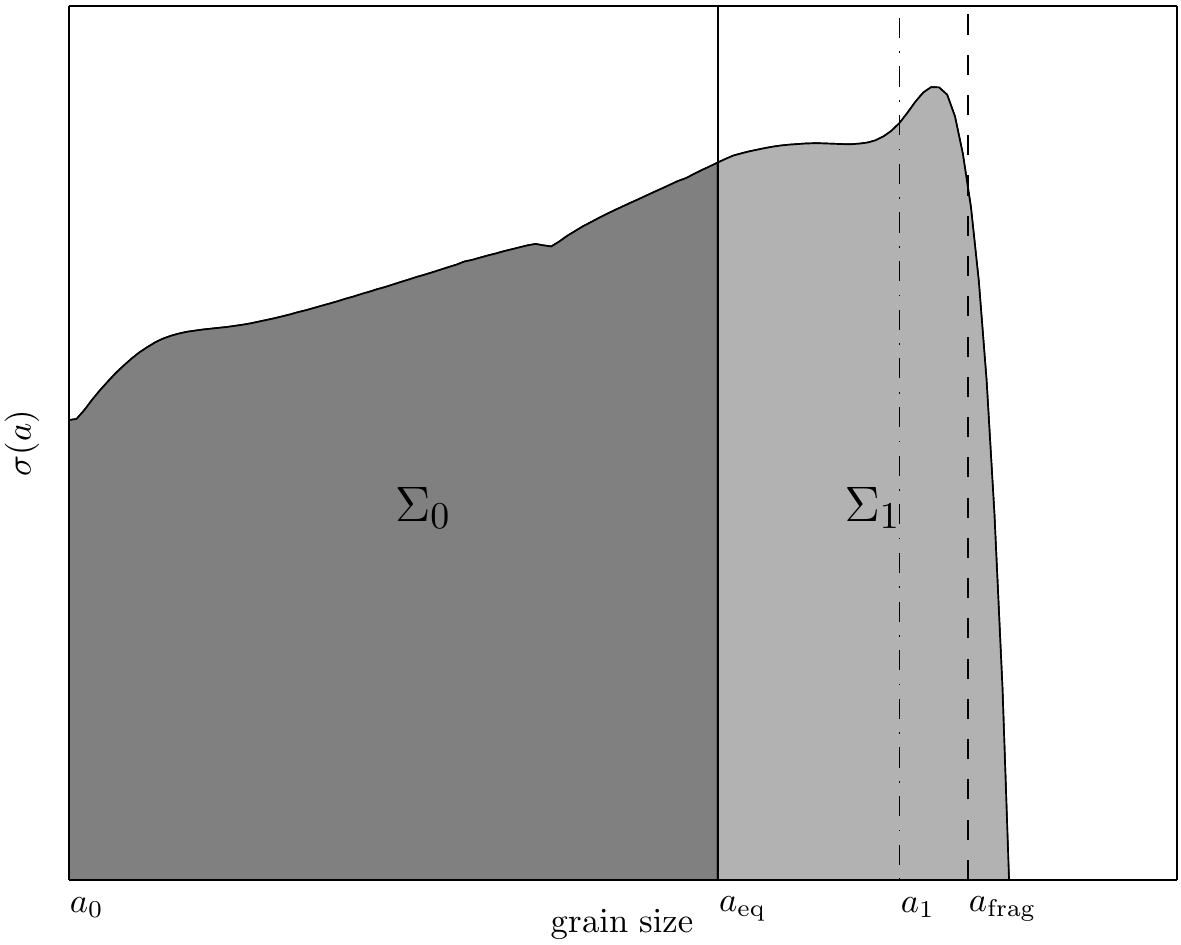}}
  \caption{Sketch of the two-population representation of the size distribution. The solid black curve is an arbitrary dust size distribution with the upper limit of $a_\mathrm{frag}$. The dark shaded area is the \emph{small population}, i.e. the part of the size distribution which is not influenced by drift velocities because particles of these sizes are low enough to be tightly coupled to the gas. The light gray area is the \emph{large population}, i.e. the grains that are significantly drifting inwards. The total mass contained in the small and large population is given by $\Sigma_0$ and $\Sigma_1$, while the representative size is $a_0$ and $a_1$, respectively.}
  \label{fig:sketch}
\end{figure}

\subsection{Size limits}\label{sec:two_pop_model:size_limits}
We crudely simplify the grain size distribution by only two representative sizes, $a_0(r)$ and $a_1(r)$, where $a_0(r)$ is the monomer size and $a_1(r)$ is the representative size of the large grain population (see also the sketch in Fig~\ref{fig:sketch}). While $a_0(r)$ is taken to be constant in time and space, $a_1(r)$  increases with time as particles grow and the disk structure evolves. This growth is not without limits. As explained in Section~\ref{sec:background}, we consider the two most important barriers towards further growth, namely grain fragmentation and radial drift. We find that $a_1(r)$ can also decrease with time if the growth limits drop to smaller sizes.

\subsubsection{Turbulent fragmentation}\label{sec:two_pop_model:size_limits:turb_frag}
In the case where the fragmentation of dust grains due to turbulent relative motion is the limiting effect, analytical models of the grain size distribution have been derived in \citet{Birnstiel:2011p13845}. In these stationary and local grain size distributions, a significant fraction of the dust mass is concentrated in the largest sizes, \emph{slightly below} the limiting Stokes number $\St_\text{frag}$. We parametrize this offset by an order-of-unity constant $f_\mathrm{f}$ such that the representative size for the largest grains in a fragmentation-dominated distribution can be stated as (using Eqns.~\ref{eq:def_stokesnumber} and \ref{eq:St_f})
\begin{equation}
a_\mathrm{frag} = f_\mathrm{f}\, \frac{2}{3 \, \pi} \, \frac{\Siggas}{\rhos \alphat} \, \frac{\uf^2}{\csound^2}.
\label{eq:a_frag}
\end{equation}

\subsubsection{Radial drift}\label{sec:two_pop_model:size_limits:rad_drift}
Radial drift can also set a limit to the local size of dust grains if the time within which particles are removed by drift is similar or less than the time scale at which the particles are replenished, i.e. the growth time scale. To derive the growth time scale
\begin{equation}
\tau_\mathrm{grow} = \frac{a}{\dot a},
\end{equation}
we follow \citet{Brauer:2008p215} by using the growth rate of monodisperse coagulation
\begin{equation}
\frac{\mathrm{d}a}{\mathrm{d}t} = \frac{\rhodust}{\rhos} \Delta u
\label{eq:a_dot}
\end{equation}
and the approximate relative velocities between the similar-sized dust grains for turbulent motion \citep[see][]{Ormel:2007p801}
\begin{equation}
\Delta u = \sqrt{3 \, \alphat \, \St} \, \csound.
\label{eq:dv_turb}
\end{equation}
Assuming $\St<1$, the Epstein drag law, and a scale height of the dust distribution of \citep{Youdin:2007p2021}
\begin{equation}
\Hd = \Hg \sqrt{\frac{\alphat}{\St}},
\label{eq:h_dust}
\end{equation} 
the growth time scale can be written as 
\begin{equation}
\tau_\mathrm{grow} \simeq \frac{1}{\Ok \, \epsilon}, 
\label{eq:tau_grow}
\end{equation}
where $\epsilon = \Sigdust/\Siggas$ is the vertically integrated dust-to-gas mass ratio and we neglected a factor of order unity. 

It should be noted that in this case (i.e. turbulent relative velocities), the growth time scale is independent of the particle size because the rate of change of the particle radius becomes directly proportional to the particle radius itself. The drift time scale
\begin{equation}
\tau_\mathrm{drift} = \frac{r}{u_\mathrm{D}}
\end{equation}
depends on the drift velocity (cf. Eq.~\ref{eq:u_drift}) and thus the Stokes number of the particle. It is given by
\begin{equation}
\tau_\mathrm{drift} = \frac{r \, V_\mathrm{k}}{\St\, \csound^2} \, \gamma^{-1},
\label{eq:tau_drift}
\end{equation}
where $V_\mathrm{k}$ is the Keplerian velocity and
\begin{equation}
\gamma = \dlnPdlnR
\label{eq:gamma}
\end{equation}
is the absolute value of the power-law index of the gas pressure profile.
By equating the growth time scale and the size dependent drift time scale, we derive an estimate of the maximum Stokes number that can be reached before radial drift removes the dust particles,
\begin{equation}
\St_\mathrm{drift} = f_\mathrm{d} \, \epsilon \, \frac{V_\mathrm{k}^2}{\csound^2} \, \gamma^{-1},
\label{eq:St_drift}
\end{equation}
or in terms of particle size (again assuming Epstein drag law),
\begin{equation}
a_\mathrm{drift} = f_\mathrm{d} \, \frac{2 \, \Sigdust}{\pi \, \rhos}\frac{V_\mathrm{k}^2}{\csound^2} \, \gamma^{-1}. 
\label{eq:a_drift}
\end{equation}
This boundary is not as sharp as the fragmentation barrier, as can be seen in Fig.~\ref{fig:sim_results_A}. We have introduced another order-of-unity factor $f_\mathrm{d}$, which we calibrate later with the detailed numerical simulations.

Strictly speaking, Eqns.~\ref{eq:dv_turb} and~\ref{eq:h_dust} are not valid for very small particles that are very tightly coupled to the gas. However this limit was derived for drifting particles, which are per definition only marginally coupled to the gas thus Eqns.~\ref{eq:dv_turb} and~\ref{eq:h_dust} are valid approximations.

\subsubsection{Fragmentation by relative drift velocities}\label{sec:two_pop_model:size_limits:drift_frag}
So far we have only considered fragmentation by turbulent relative velocities. However in the case of very low turbulent velocities, one might expect that the relative drift speed of the particles, either in the vertical or in the radial dimension, is still high enough to cause fragmentation. \change{}{Dust grains in the tenuous disk atmosphere settle towards the mid-plane with velocities that can exceed the radial ones \citep{Nakagawa:1986p2048}. In a turbulent disk, mixing counteracts this downward motion and an equilibrium state is reached on time scales less than $10^5$ years, which means that the vertical motion is not significant throughout the lifetime of the disk. In contrast to this, radial drift motion does not vanish towards the mid-plane and is therefore a systematic velocity that stays relevant throughout the lifetime of the disk. We therefore only focus on the fragmenting effects of the radial dust motion.} We show in Section~\ref{sec:discussion:drift_velocities} that this effect is irrelevant for the simulations of a turbulent disk. For laminar disks, this effect can become relevant.

To derive a size limit for this case, we need an estimate of the relative velocities induced by radial drift. From Eq.~\ref{eq:u_drift}, we can derive the relative drift speed between two particles for Stokes numbers below unity
\begin{equation}
\Delta u_\mathrm{drift} = \frac{\csound^2}{V_\mathrm{k}} \, \gamma \, \left| \St_1 - \St_2 \right|.
\label{eq:dv_drift}
\end{equation}
Expressing the Stokes number of the smaller particle as $\St_2 = N \cdot \St_1$, and equating Eq.~\ref{eq:dv_drift} with the fragmentation threshold velocity \uf, we derive
\begin{equation}
\St_\mathrm{df} = \frac{\uf \, V_\mathrm{k}}{\gamma \, \csound^2 \left(1-N\right)},
\label{eq:St_df}
\end{equation} 
which is the largest Stokes number that can be reached before the relative drift speed exceeds the fragmentation threshold. The most important collision types determining the mass gain/loss of the largest bodies is typically collisions with similar sized bodies. $N=0.5$ is therefore a reasonable assumption.

\subsection{Mass fractions \& transport equation}\label{sec:two_pop_model:mass_fractions}
In order to represent the size distribution of the dust with only two representative sizes, we also need to know the ratio of mass contained in large and small particles. We represent the mass ratio between the two populations by another yet unknown factor $f_\mathrm{m}$,
\begin{align}
\Sigma_1(r) &=  \Sigdust(r) \cdot f_\mathrm{m}(r)     \\ 
\Sigma_0(r) &=  \Sigdust(r) \cdot (1-f_\mathrm{m}(r)),
\label{eq:sig_total}
\end{align}
where $\Sigma_0$ is the surface density in the small grains while $\Sigma_1$ is the surface density in large grains. This representation is schematically shown in Fig.~\ref{fig:sketch}.
In the case of a dust size distribution which is in coagulation/fragmentation equilibrium, small dust is replenished. In the case of a drift-limited size distribution, fragmentation is not efficient enough since particles drift inwards before reaching sizes at which they fragment.

For particles with a Stokes number below unity, this allows us to evolve the dust distribution by a single advection-diffusion equation
\begin{equation}
\frac{\del \Sigdust}{\del t} + \frac{1}{r} \frac{\del}{\del r} \left[ r \left(\Sigdust \, \bar u - D_\mathrm{gas} \, \Siggas \, \frac{\del}{\del r}\left( \frac{\Sigdust}{\Siggas} \right) \right) \right] = 0,
\label{eq:dust_simple}
\end{equation}
using the mass weighted velocity
\begin{equation}
\bar u = (1-f_\mathrm{m}(r)) \cdot u_0 + f_\mathrm{m}(r) \cdot u_1,
\label{eq:u_bar}
\end{equation}
and the gas diffusivity $D_\mathrm{gas}$. Here, $u_0$ and $u_1$ denote the radial velocity of the dust sizes $a_0$ and $a_1$, respectively. The radial velocity of dust grains is the sum of the drift velocity given by Eq.~\ref{eq:u_drift} and the gas drag induced velocity
\begin{equation}
u_\mathrm{drag} = \frac{1}{1 + \St^2} \cdot u_\mathrm{gas},
\end{equation}
where $u_\mathrm{gas}$ is the gas radial velocity.
A more thorough justification of Eq.~\ref{eq:dust_simple} is given in Appendix~\ref{app:simplified_equation}.

\begin{figure}[tb]
  \centering
  \resizebox{\hsize}{!}{\includegraphics{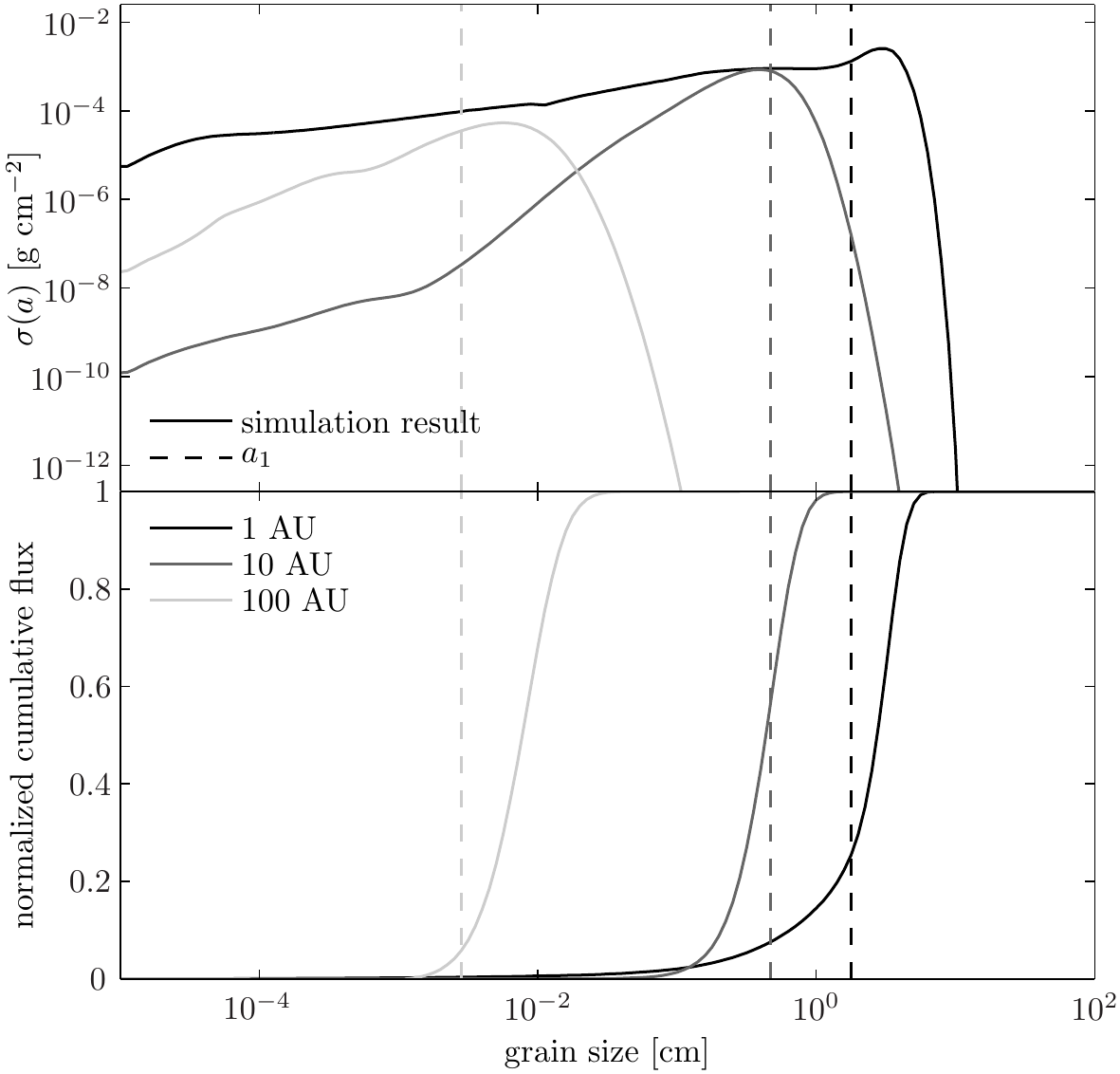}}
  \caption{Grain size distributions (top panel) and the cumulative flux (bottom panel) for the simulation result which is displayed in the right panel of Fig.~\ref{fig:sim_results_A} at 1, 10, and 100~AU. The different radii are represented by different gray scale colors and the dashed lines mark the representative grain size $a_1$.}
  \label{fig:mass_distributions}
\end{figure}

\begin{figure}[tb]
  \centering
  \resizebox{0.99\hsize}{!}{\includegraphics{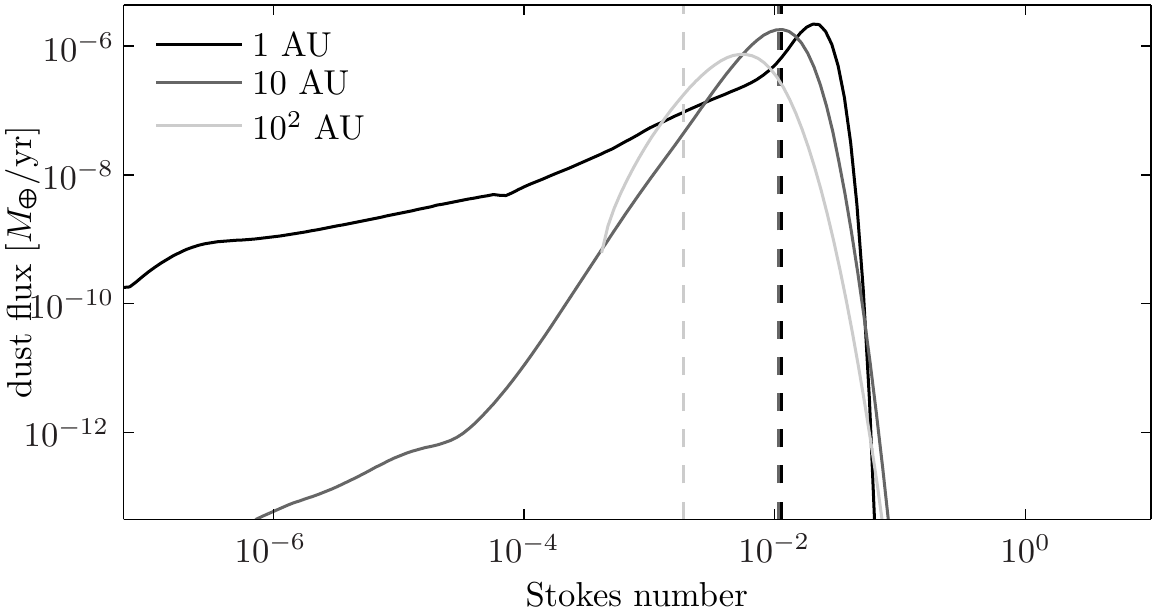}}
  \caption{The dust flux distribution $\sigma(a) \cdot u(a) $ as function of Stokes number for the same dust distributions as shown in Fig.~\ref{fig:mass_distributions}.}
  \label{fig:flux_distributions}
\end{figure}

\subsection{Particle growth}\label{sec:two_pop_model:particle_growth}
The mass flux and thus the whole dust evolution is strongly influenced by the growth time scale. This is due to the fact that the dust-to-gas coupling and thus the radial velocity strongly depends on the particle size. Thus, particles in the outer parts, which grow less quickly, start to drift at later times, once they have reached sizes large enough to be influenced by drift. 
From the growth time scale (i.e. the size-doubling time) given in Eq.~\ref{eq:tau_grow} we can estimate the time it takes to grow from the smallest size $a_0$ to a maximum size $a_1$
\begin{equation}
t_\mathrm{grow} = \tau_\mathrm{grow} \cdot \ln\left(\frac{a_1}{a_0}\right),
\label{eq:t_grow}
\end{equation}
where
\begin{equation}
a_1 = \min\left(a_\mathrm{drift},a_\mathrm{frag}\right).
\end{equation}
In order to take this \emph{delayed drift} into account, we define a time dependence of the larger representative size
\begin{equation}
a_1(t) = \min \left[ a_1 , a_0 \cdot \exp\left(\frac{t}{\tau_\mathrm{grow}}\right)\right].
\label{eq:a_1_of_t}
\end{equation}
This is an oversimplified estimate since different growth mechanisms are at play at different sizes, however we found a good agreement between our estimate and the simulation result. This can be seen in Figs.~\ref{fig:sim_results_A} and \ref{fig:sim_results_B}, where the dash-dotted curve represents the time dependent upper size limit according to Eq.~\ref{eq:a_1_of_t}. Only at early times (left panels of Figs.~\ref{fig:sim_results_A} and \ref{fig:sim_results_B}), $a_1(t)$ differs from the fragmentation or drift barrier.

It can also be seen that this estimate is slightly too low, i.e. it is slightly below the upper size of the simulated dust distribution at all radii $\gtrsim 20$~AU in the left panel of Fig.~\ref{fig:sim_results_B}. This is due to the fact that the growth time scale of Eq.~\ref{eq:tau_grow} implicitly assumes growth to be driven only by equal sized collisions from turbulent motion according to Eq.~\ref{eq:dv_turb}. However, the initial stages of particle growth involve different turbulent regimes \citep[see][]{Ormel:2007p801}, which lead to faster growth for higher values of \alphat.

\begin{table}
\caption{Best-fit model parameters}
\label{tab:parameters}
\centering
\begin{tabular}{c | c c}
\hline\hline
parameter      & drift limited & fragmentation\\
               & case          & limited case\\
\hline
$f_\mathrm{f}$ & --            & 0.37 \\
$f_\mathrm{d}$ & 0.55          & --   \\
$f_\mathrm{m}$ & 0.97          & 0.75 \\
\hline
\end{tabular}
\tablefoot{The values in this table represent the values of the model parameters that best reproduce the results of the full simulations.}
\end{table}

\subsection{Calibration and test cases}\label{sec:two_pop_model:calib_test_cases}
In the semi-analytic model described above, we have introduced three factors $f_\mathrm{f}$, $f_\mathrm{d}$, and $f_\mathrm{m}$. These factors are necessary since our model is not derived in a rigorously analytical way, but includes several assumptions and estimates. In order to still achieve a good agreement with the detailed numerical simulations, we need to calibrate these order-of-unity factors by comparison to a grid of the full simulation results.

In order to do this, we divide the grain size distribution at each radius in two populations: the \emph{large population}, which are all grains whose drift velocity is higher than the gas drag induced velocity, while everything that is smaller belongs to the \emph{small population}. The gas radial velocity can be written as \citep[see][]{Takeuchi:2002p3167}
\begin{equation}
u_\mathrm{g} = -\frac{3 \, \alphat \, \csound^2}{V_\mathrm{k}} \,  \left( 2 - p - q \right),
\label{eq:u_gas}
\end{equation}
where $p$ is defined as $\Siggas\propto r^{-p}$ and $q$ as $T\propto r^{-q}$, and \alphat is taken to be constant. Thus, the gas drag velocity is lower than the drift velocity for particles with a Stokes number larger than
\begin{equation}
\St_\mathrm{eq} \gtrsim \frac{3 \, \left|2-p-q\right|}{\left|\gamma\right|}\, \alphat,
\label{eq:St_eq}
\end{equation}
which for typical values ($p=1$, $q=0.5$, $\gamma=-p-\frac{q}{2}-\frac{3}{2}$) yields $\St_\mathrm{eq}\gtrsim \alphat/2$.

As described in Section~\ref{sec:two_pop_model:size_limits}, the fragmentation is a rather strict upper limit to the size distribution. The dust mass flow is therefore not dominated by grain sizes at the fragmentation barrier but slightly below it. We therefore calculated a flux-averaged grain size for the large population
\begin{equation}
\bar a(r) = \int_{a_\mathrm{eq}}^\infty \sigma(a,r) \, u(a) \, \mathrm{d}a \Bigg/ 
\int_{a_\mathrm{eq}}^\infty \frac{\sigma(a,r) \, u(a) } {a} \, \mathrm{d}a,
\label{eq:a_bar}
\end{equation}
where
\begin{equation}
a_\mathrm{eq} = \St_\mathrm{eq} \, \frac{2\,\Siggas}{\pi \, \rhos}
\label{eq:a_eq}
\end{equation}
is the grain size corresponding to a Stokes number of $\St_\mathrm{eq}$ and $u(a)$ is the total radial velocity of grain size $a$. \change{}{We compared the outcome of the simplified model to a grid of 39 full simulations.}\footnote{\change{}{We carried out 39 simulations using as initial condition either self-similar solutions (Eq.~\ref{eq:sig_gas}) with $r_c = 20, 60, 200$~AU, $\alphat = 10^{-2}, 10^{-3}, 10^{-5}$, disk masses of 0.01 and 0.1 $M_\odot$, fragmentation velocities of 1, 3, and 10~m~s$^{-1}$ or power-law models with gas surface density exponents of -1 with outer radii of 250~AU.}} Good agreement of this flux averaged size with $a_\mathrm{frag}$ was found by tuning $f_\mathrm{f}$ to a value of 0.37.

\begin{figure}[tb]
  \centering
  \resizebox{\hsize}{!}{\includegraphics{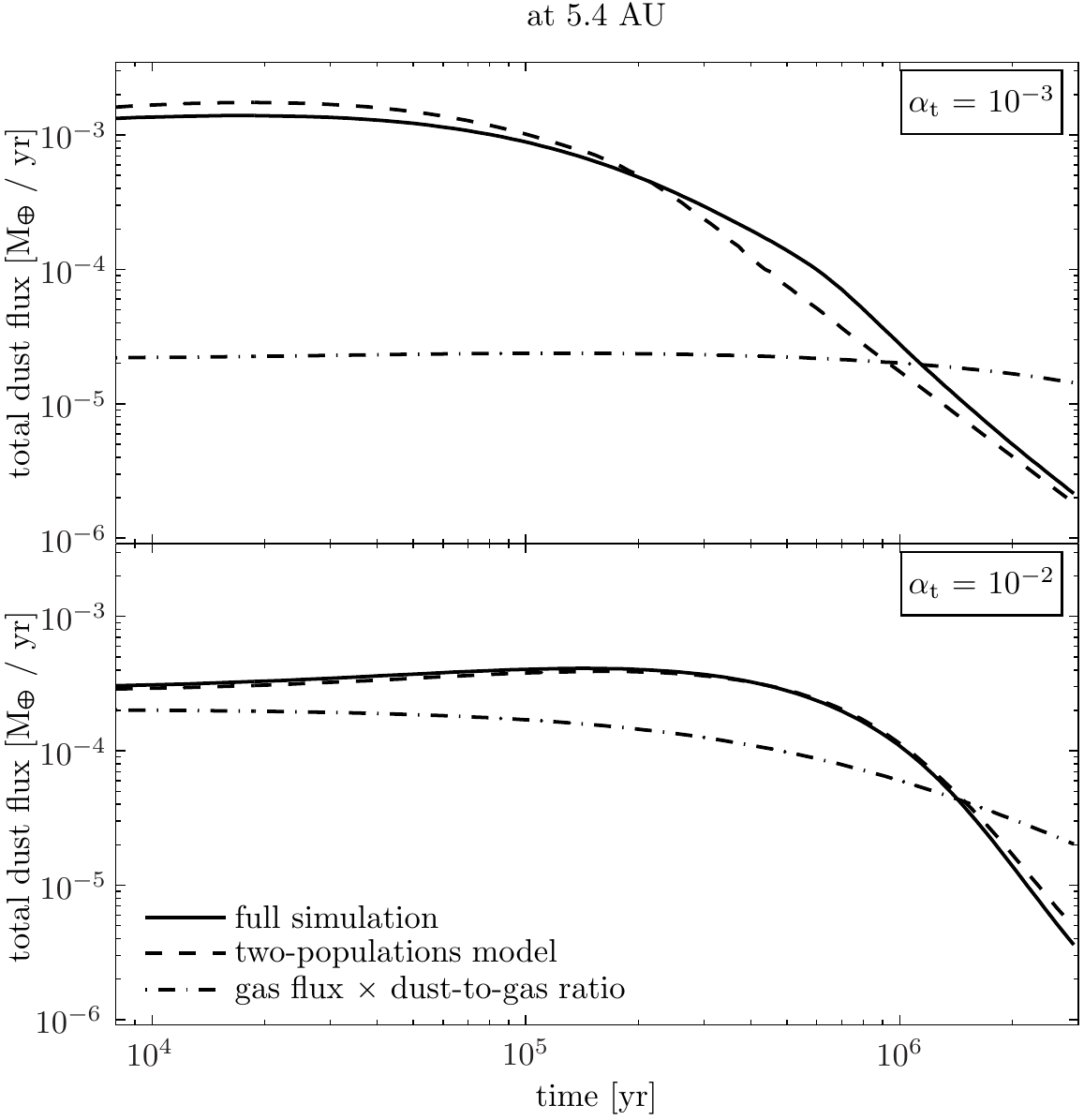}}
  \caption{Total inward mass flux of dust for the full simulations (solid line), for the two-population model (dashed line) and the scaled gas accretion rate (dot-dashed line) in earth masses per year at 5.4~AU. The upper and lower panel correspond to the weak and strong turbulence simulations, respectively.}
  \label{fig:dust_flux_t}
\end{figure}

\begin{figure*}[tb]
  \centering
  \resizebox{!}{0.48\hsize}{\includegraphics{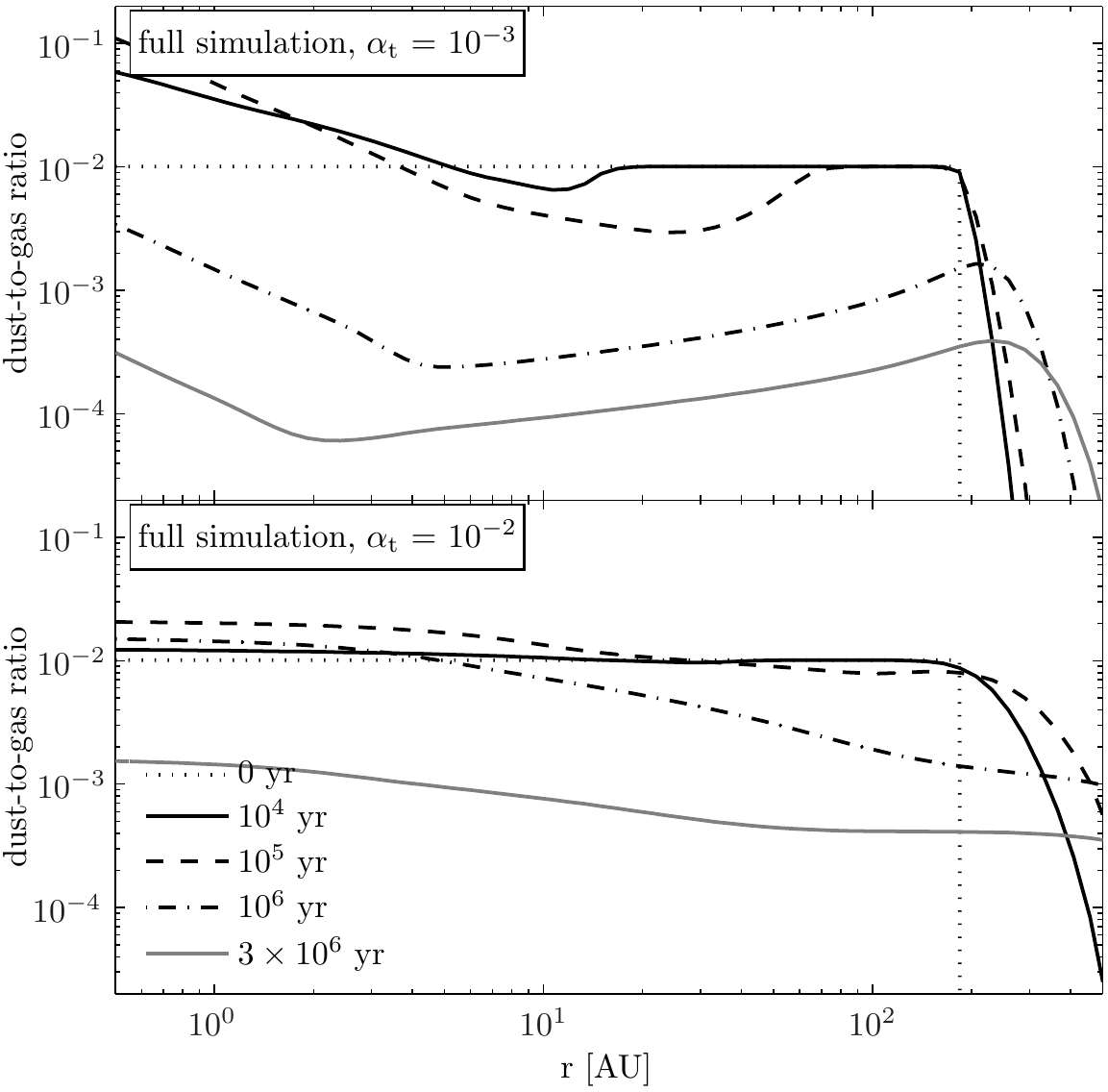}}
  \resizebox{!}{0.48\hsize}{\includegraphics{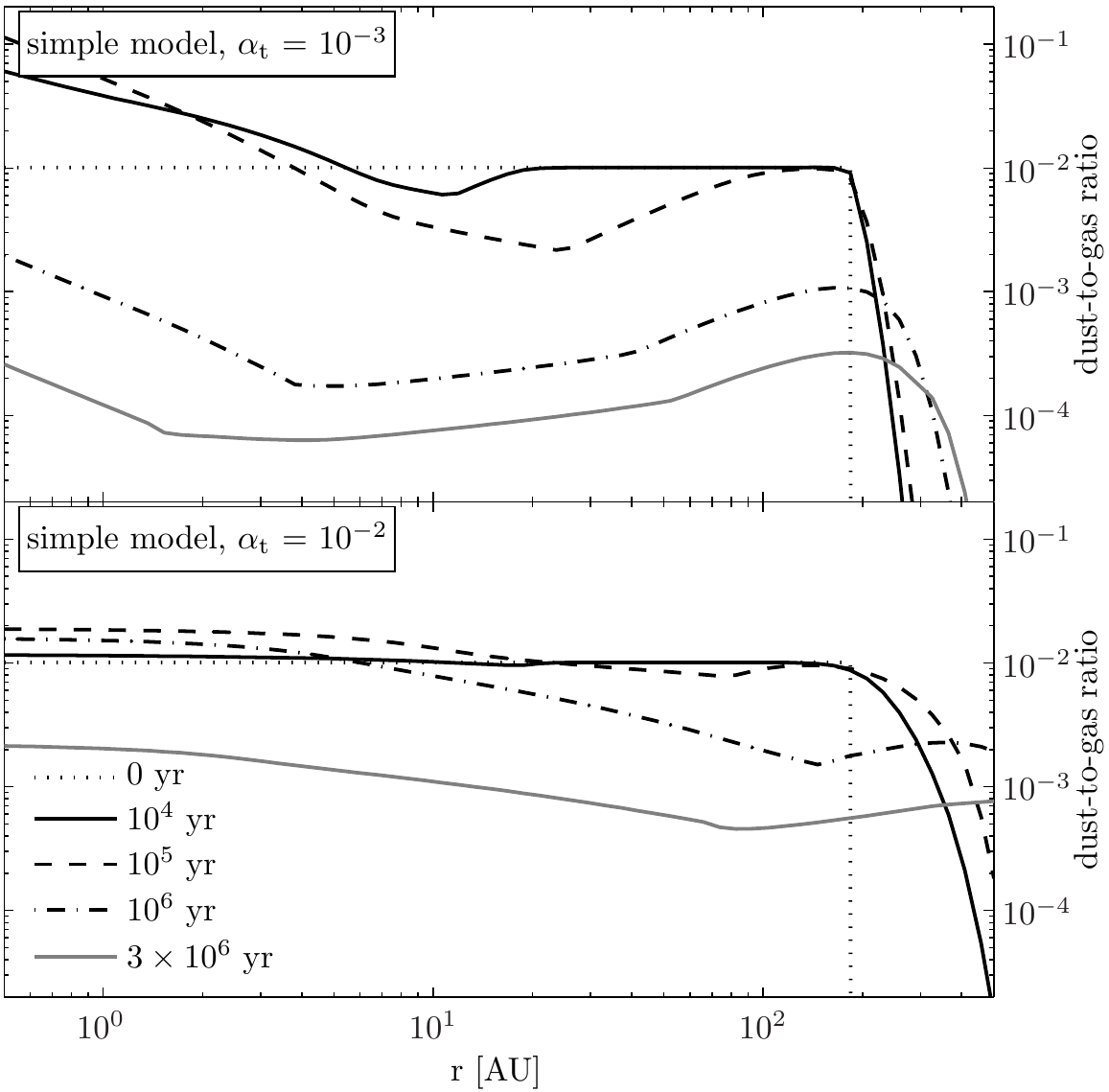}}
  \caption{Evolution of the vertically integrated dust-to-gas ratio for the weak and the strong turbulence simulation (upper and lower panels respectively). The panels on the left hand side show the results of the full simulation, the panels on the right hand side those of the simplified model. The dotted line is the initial condition which is taken to be a constant dust-to-gas ratio of 0.01 up to 180~AU. Even the smallest grains at the low gas densities beyond 180~AU would decouple and quickly drift inside. We chose to remove the dust beyond 180~AU to avoid a wave of dust sweeping inwards, which would be caused by an initial condition with constant dust-to-gas ratio throughout the disk.}
  \label{fig:dust_to_gas}
\end{figure*}

For the case of a drift limited size distribution, the size limit is not as strict as in the fragmentation limited case, i.e. there can exist grains slightly larger than $a_\mathrm{drift}$. The same method as above yields therefore a value of $f_\mathrm{d}$ of 0.55, which is higher than in the case of a fragmentation limited size distribution. \change{}{These values of $f_\mathrm{f}$ and $f_\mathrm{d}$ should thus be universal, as long as the collision outcome model is not changed.}

The dashed and solid lines in Fig.~\ref{fig:sim_results_A} denote the representative grain size for a drift limited and a fragmentation limited maximum grain size, respectively. It can be seen that these sizes, together with the time dependence introduced in Section~\ref{sec:two_pop_model:particle_growth}, describe the limits of the grain size distribution well.

In order to properly capture the radial transport of dust with this simple model, we need also to calibrate the parameter $f_\mathrm{m}$, which describes how the mass is distributed between the two populations. Typical size distributions can be seen in the top panel of Fig.~\ref{fig:mass_distributions}, denoted by the solid lines. Both the drift and the fragmentation limited distributions have most of the mass in the large grain population. However fragmentation limited distributions (e.g., the solid black line in the top panel of Fig.~\ref{fig:mass_distributions}) have a larger fraction of mass in the small sizes compared to drift limited distributions (solid gray lines). The value of $f_\mathrm{m}$ thus depends on whether drift or fragmentation is the limiting factor. We found typical values of $f_\mathrm{m}$ around 0.75 for the fragmentation limited case and around 0.97 for the drift limited case. All values are summarized in Table~\ref{tab:parameters}.

It should be noted that the flux is typically determined by the large population. This is shown in Fig.~\ref{fig:flux_distributions}. The plotted quantity $\sigma(a) \cdot u(a)$ is defined such that integration over $\log(a)$ (or equally $\log(\St)$) yields the total dust flux. This means that a one-population model, which neglects the small population can describe the dust evolution in the drift limited case well. However in the case of a fragmentation-dominated distribution or an outward-spreading gas disk, the small population can contain enough mass or have a significant velocity to be relevant and generally better agreement is obtained if this population is taken into account, while the additional complications and computational costs of the two-population model are minimal.

Another important test case is whether the total dust mass flux is reproduced. In Fig.~\ref{fig:dust_flux_t} we compare the total dust flux from the full simulations (solid line) to our two-population model (dashed line).
The total dust flux \change{from the full simulations}{} is calculated as
\begin{equation}
\dot M_\mathrm{d}(r) = 2\,\pi\,r\,\int_0^\infty \sigma(a,r)\, u(a)\, \mathrm{dln}a. 
\label{eq:total_simulation_flux}
\end{equation}
The dash-dotted line \change{}{in} Fig.~\ref{fig:dust_flux_t} represents the gas accretion rate multiplied by the initial dust-to-gas ratio, i.e. the dust mass flux that would be found if all dust particles were perfectly coupled to the gas accretion flow. The error of the toy model is always within a factor of two, which is a surprisingly good result considering the simplicity of the model. \highlight{The results are roughly consistent with the order-of-magnitude estimates of \citet{Klahr:2006p7719}.}

Finally, Fig.~\ref{fig:dust_to_gas} compares the profiles of the dust-to-gas ratio. It can be seen that the two-population model reproduces the full simulation results reasonabl well.

\section{Discussion}\label{sec:discussion}
The two-population model presented in Section~\ref{sec:two_pop_model} was shown to reproduce the time evolution of several important quantities, such as the dust-to-gas ratio, the dust mass flux and also the size of the drifting particles. In this section we discuss some of the applications of this model.

\subsection{Fragmentation by drift velocities}\label{sec:discussion:drift_velocities}
In the following subsection we investigate whether the relative velocities due to radial drift of the particles can become high enough to play a role. For this to be the case, the size limit of Eq.~\ref{eq:St_df} must be smaller than the limits set by both turbulent fragmentation (cf., Eq.~\ref{eq:St_f}) and by transport (cf., Eq.~\ref{eq:St_drift}). These two condition can be rewritten as
\begin{equation}
\alphat < \frac{\uf}{V_\mathrm{k}} \frac{\gamma \, \left(1-N\right)}{3}
\label{eq:condition_1}
\end{equation} 
and
\begin{equation}
\epsilon > \frac{\uf}{V_\mathrm{k}} \, \frac{1}{f_\mathrm{d} \, \left(1-N\right)},
\label{eq:condition_2}
\end{equation}
respectively.
Considering a typical disk where $\gamma \simeq 2.75$ and assuming $N=0.5$, we find that for strong turbulence ($\alphat>10^{-2}$), the first condition is never fulfilled even for high fragmentation velocities of $\sim 10$ m s$^{-1}$. For weaker turbulence such as $\alphat\lesssim 10^{-3}$, the regions beyond $\sim 10$~AU can fulfill Eq.~\ref{eq:condition_1}.
Eq.~\ref{eq:condition_2} sets constraints from the opposite direction, such that the region between about 10 and 60~AU could be influenced by drift induced fragmentation.

To quantify the error we do by considering only the size limits by drift and fragmentation, we consider the ratio between $\St_\mathrm{drift}$ and $\St_\mathrm{df}$, which is given by
\begin{equation}
\frac{\St_\mathrm{drift}}{\St_\mathrm{df}} \simeq 2.5 \cdot \left(\frac{\epsilon}{0.01}\right) \, \left(\frac{\uf}{10 \, \mathrm{m~s}^{-1}}\right)^{-1} \, \left(\frac{r}{10\mathrm{~AU}}\right)^{-\frac{1}{2}}.
\label{eq:St_ratio}
\end{equation}
It is important to note that this deviation only needs to be considered in regions where it is larger than unity and turbulent fragmentation does not play a role. Since the regions further inside are dominated by turbulent fragmentation, the largest error we do is overestimating the upper end of the distribution by a factor of about 2.5. As we go to 60~AU the error is decreasing while the regions beyond 60~AU are in any case limited by particle drift and not by fragmentation. Decreasing the fragmentation threshold velocity seems to increase the error, however in this case the region where drift induced fragmentation applies, then disappears because turbulent fragmentation becomes more important.

Most importantly, we see that Eq.~\ref{eq:St_ratio} depends on the dust-to-gas ratio. Since the region beyond 10~AU is strongly drifting, the dust-to-gas ratio $\epsilon$ is quickly decreasing as can be seen in Fig.~\ref{fig:dust_to_gas}. Therefore, we can safely ignore the size limit set by Eq.~\ref{eq:St_df} if we are only concerned about the evolution of the dust surface density or the upper end of the dust size distribution.

We also carried out simulations with a very small amount of turbulence ($\alphat=10^{-5}$). As expected, particles grow to large Stokes numbers until fragmentation due to radial drift sets in. However even in this case, the dust-to-gas ratio drops so quickly that after 1~Myr the drift limit (Eq.~\ref{eq:St_drift}) becomes everywhere smaller than the limit set by drift induced fragmentation (Eq.~\ref{eq:St_df}), such that grains do not fragment anymore. In terms of the presence of small dust, this effect does matter: it means that a drift-limited size distribution cannot sustain efficient fragmentation on long timescales and thus contains much smaller amounts of small dust grains (see right panel of Fig.~\ref{fig:sim_results_A}). The observational fact that disks are observed to be rich in small dust for several millions of years thus seems to be in contradiction with a completely drift-dominated disk unless there is an external source of small dust \citep[see][]{Dominik:2008p4626} or small dust is mixed and fragmented into surface layers with stronger turbulence, an effect that we cannot take into account in a vertically integrated dust evolution code. We have shown that radial drift alone cannot sustain the relative velocities to keep fragmentation active for the observed lifetimes of protoplanetary disks. Fragmentation together with vertical mixing is therefore likely driven by turbulent motion.

\subsection{Analytical dust surface density profiles}\label{sec:discussion:surfdensprofile}
The evolution of the dust-to-gas ratio is important for theories of planetesimal formation \citep[e.g.,][]{Johansen:2007p4788,Johansen:2011p14449} and planet formation (see \highlight{\citealp{Klahr:2006p7719}}; \citealp{Lissauer:2007p15631}; \citealp{Mordasini:2009p12261}). Also a constant dust-to-gas ratio is usually assumed for deriving disk masses from dust continuum observations \citep[e.g.,][]{Andrews:2005p3779,Andrews:2007p3380}, which is one of the biggest sources of error in disk mass estimates. In this section, we investigate how the dust surface density and thus the dust-to-gas ratio changes with time for a typical model of a circumstellar disk, starting with a constant dust-to-gas ratio of 0.01. The following discussion is based on two important aspects:
\begin{itemize}
\item 
Most of the (dust) mass resides in the outer regions of the disk. Due to the long evolution time scale at these radii, these regions provide a mass reservoir for the inner parts of the disk, as was also found in \citet{Kornet:2004p6299} or \citet{Garaud:2007p405}.
\item 
As discussed in previous sections, the dust mass is concentrated towards the upper end of the size spectrum, which is also where the drift velocity is the highest. This means that the dust flux is governed by the upper end of the size distribution.
\end{itemize}

A semi-steady dust surface density is then set by the rate at which the outer regions provide dust which is inward drifting with a velocity of $u$. Mass conservation then dictates that the dust accretion rate
\begin{equation}
\dot M_\mathrm{d} = 2\, \pi \, r \, \Sigdust \, u
\end{equation}
has to be constant for all $r$. This yields a dust surface density profile
\begin{equation}
\Sigdust = \frac{\dot M_\mathrm{d}}{2 \pi} \frac{1}{r \, u}.
\end{equation}
For a drift-dominated distribution, we can derive the drift velocity of the representative size $a_1(t)$ from Eqns.~\ref{eq:St_drift} and \ref{eq:u_drift}, which results in a surface density profile given by
\begin{equation}
  \begin{split}
\Sigma_\mathrm{d,drift} &= \sqrt{\frac{\dot M_\mathrm{d}}{2\,\pi\,f_\mathrm{d}}\frac{\Siggas}{r^2 \, \Ok}} \\
&\propto \sqrt{\frac{\Siggas}{r^2 \, \Ok}},
\label{eq:sigd_drift}
  \end{split}
\end{equation}
which for a gas surface density profile with index $p=1$ is proportional to $r^{-3/4}$.

For the case of a fragmentation-dominated distribution, such a result cannot be obtained as uniquely, because very small particles which are smaller than $a_\mathrm{eq}$ (i.e., they have a Stokes number of $\St \lesssim \alphat/2$, see Eqns.~\ref{eq:St_eq} and \ref{eq:a_eq}) are coupled to the gas, while for somewhat larger particles, radial drift starts to play a role. For a dust surface density distribution with an upper size limit larger than $a_\mathrm{eq}$ (cf. Eq.~\ref{eq:a_eq}), we can derive a stationary surface density profile as we did before but using Eq.~\ref{eq:a_frag} as upper size limit. This yields
\begin{equation}
\begin{split}
\Sigma_\mathrm{d,frag}  &= \frac{3 \dot M_\mathrm{d}}{2\, \pi} \, \frac{\alphat}{f_\mathrm{f}\, \gamma \, \uf^2}\,\Ok\\
                        &\propto \frac{\alphat \, \Ok}{\uf^2 \, \gamma},
\label{eq:sigd_frag}
\end{split}
\end{equation}
which for constant values of \alphat and \uf is proportional to $r^{-1.5}$, as the Minimum Mass Solar Nebula profile \citep{Weidenschilling:1977p15694,Hayashi:1981p15696}. 

\begin{figure}[htb]
  \centering
  \resizebox{\hsize}{!}{\includegraphics{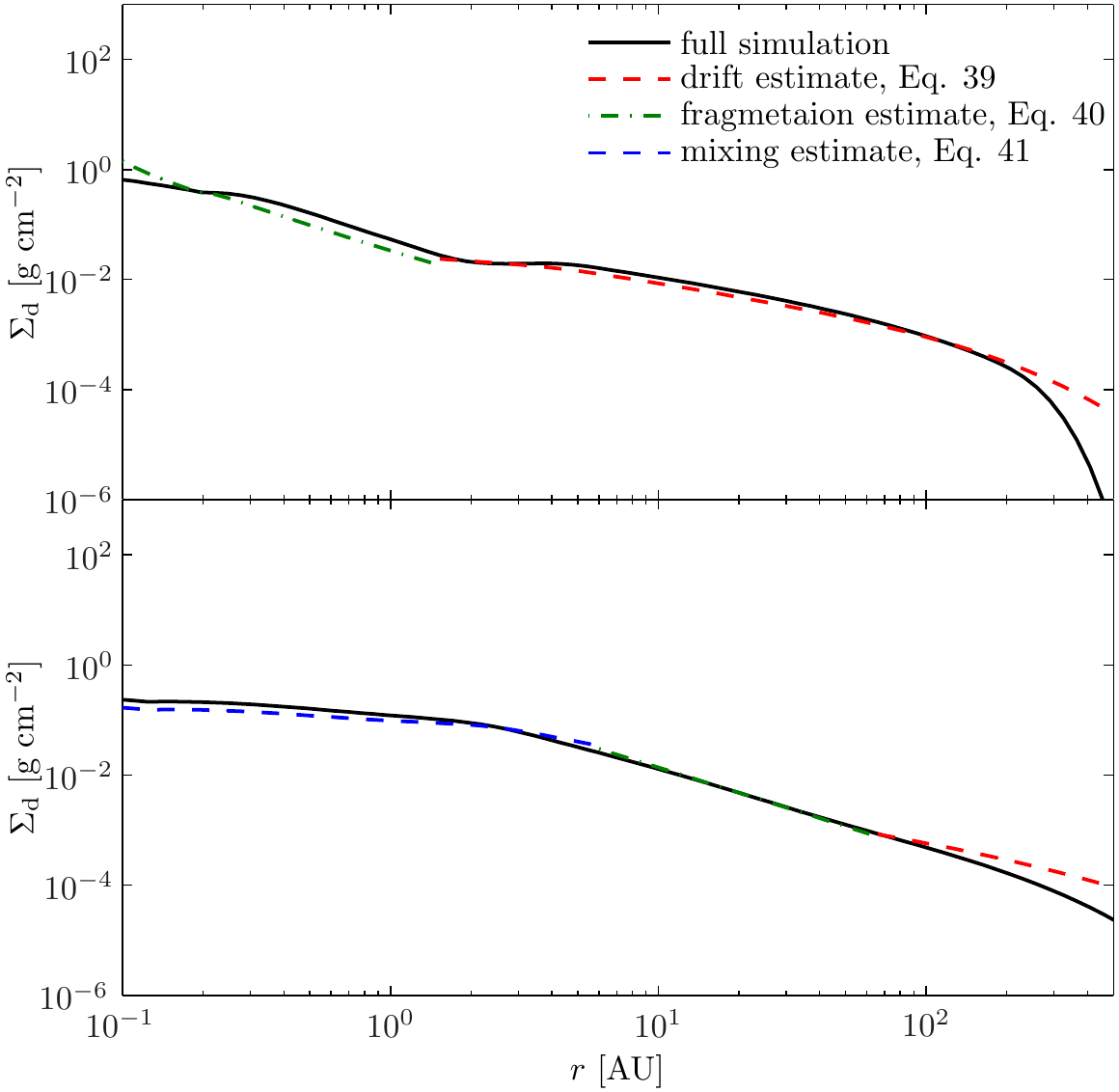}}
  \caption{Comparison between simulation results and the simple estimates derived in Section~\ref{sec:discussion:surfdensprofile} after 3~Myrs of dust evolution. The upper panel is for a turbulence value of $\alphat = 10^{-3}$, the lower one for $\alphat=10^{-2}$. The estimates have been calculated using the dust accretion rate of the simulations at 50~AU which are $1.6\times 10^{-6}$ and $2.4\times 10^{-6} M_\Earth$/yr, respectively. The division between the different fit formulas is described in Section~\ref{sec:discussion:surfdensprofile}.}
  \label{fig:dust_profiles}
\end{figure}

A dust distribution that consists entirely of small dust grains ($\St \lesssim \alphat/2$ for the largest grains) should be so well coupled to the gas that the dust-to-gas ratio is constant, i.e.
\begin{equation}
\begin{split}
\Sigma_\mathrm{d,mixed} &= \frac{\dot M_\mathrm{d}}{2\,\pi} \frac{\Omega}{3\,\alphat\, \csound^2} \frac{1}{\left|2-p-q\right|}\\
                        &\propto \Siggas,
\label{eq:sigd_mixed}
\end{split}
\end{equation}
because the dust is just co-moving with the gas flow.

We have now derived three different shapes of the dust surface density distribution. In Fig.~\ref{fig:dust_profiles}, we compare our simple estimates to the previously discussed simulation results of the full model. Both simulations are for a 0.1~${M}_\odot$ disk with $r_c=60$~AU around a solar mass star assuming a fragmentation velocity of 10~m~s$^{-1}$. The turbulence in the simulation of the upper panel is weaker ($\alphat = 10^{-3}$) than in the bottom panel ($\alphat = 10^{-2}$).
It can be seen that the simulation in the upper panel becomes drift-dominated in the outer parts (i.e. $a_\mathrm{drift}<a_\mathrm{frag}$). The drift estimate (cf., Eq.~\ref{eq:sigd_drift}) agrees well with the simulation result in the regions between about 2 and 200~AU, while the inner parts follow the fragmentation limited estimate, Eq.~\ref{eq:sigd_frag}.
The estimates do not capture the drop-off in the outermost parts of the disks because they assume a constant accretion rate, while in reality, the surface density and thus the accretion rate must decline towards the outer end of the disk.
The stronger turbulence in the bottom panel causes the upper limit of the dust size distribution to decrease. The fragmentation-dominated region now reaches out to larger radii at all times while grains in the innermost regions become so small that they are well coupled to the gas and thus can be described by the mixing estimate of Eq.~\ref{eq:sigd_mixed}. Only the regions beyond about 60~AU stay drift-dominated. For a lower fragmentation threshold velocity of 1~m~s$^{-1}$, the entire disk was found to be fragmentation or mixing dominated. 

The results shown in Fig.~\ref{fig:dust_profiles} are for 3~Myrs of dust evolution. At earlier times, for example 1~Myr (not shown), the dust-to-gas ratio in the outer parts is still higher and thus the drift limited region is smaller, covering only the region beyond 4 and 150~AU for the low and high turbulence case, respectively.

\begin{figure}[htb]
  \centering
  \resizebox{\hsize}{!}{\includegraphics{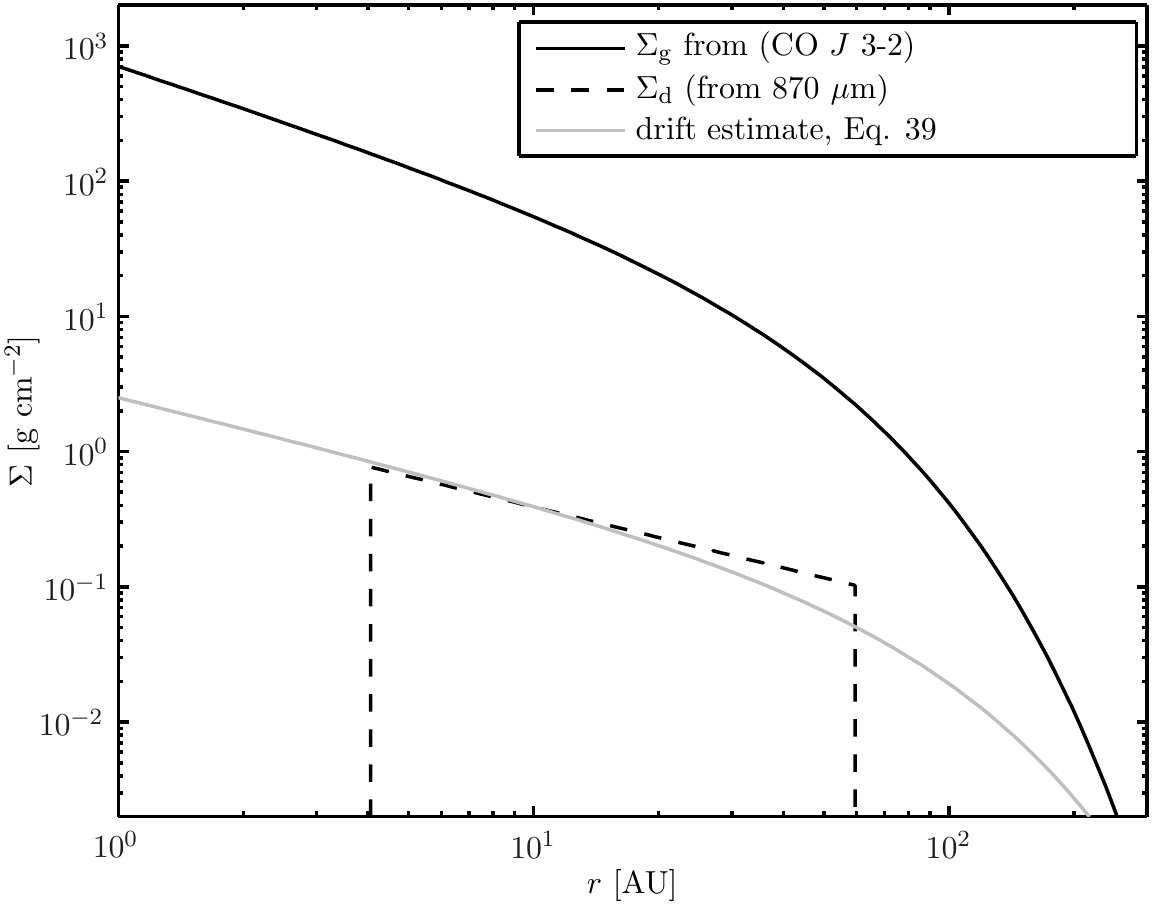}}
  \caption{\change{}{The best-fit models of \citet{Andrews:2012p16676} for the dust and gas profiles of TW Hya (solid and dashed black lines, respectively) and the analytical, drift limited dust profile as in Eq.~\ref{eq:sigd_drift} (solid grey line), using the gas profile of \citet{Andrews:2012p16676} is used as input.}}
  \label{fig:comparison_andrews2012}
\end{figure}

\change{}{\citet{Andrews:2012p16676} have recently found a discrepancy in the sizes and shapes of the dust and gas surface densities in the disk of TW Hya, based on observations of 870~$\mu$m dust emission and CO $J$ 3-2 emission. In Fig.~\ref{fig:comparison_andrews2012}, we plot these observationally derived profiles along with the analytical, drift limited dust surface density (using the observational gas surface density in Eq.~\ref{eq:sigd_drift}). The analytical profile has been scaled to fit the value of the dust profile from \citet{Andrews:2012p16676} at 10~AU. Such a profile would be expected for a moderate level of turbulence at late stages of disk evolution (the age of TW Hya is estimated to be about 10~Myrs, see \citealp{Kastner:1997p16679,Webb:1999p16681}). Fig.~\ref{fig:comparison_andrews2012} shows that the shape of the analytical solution agrees very well with the observations. One notable difference is the shape of the outer edge. \citet{Andrews:2012p16676} note that the remnant mismatch between their best-fit model and the 870 $\mu$m emission is likely related to the shape of the outer edge, which in their case is a sharp cut-off. In contrast to that, our model assumes a constant dust mass accretion rate through the disk. As stated earlier, the mass accretion rate has to drop off at some radius as the dust reservoir in the outer regions gets depleted. The two dust profile in Fig.~\ref{fig:comparison_andrews2012} thus most probably represent two extreme cases, i.e. a too sharp and a too smooth outer edge.}

\subsection{Evolution of the dust-to-gas ratio and dust accretion rate}\label{sec:discussion:d2g_evolution}
The results presented in the previous section can directly be translated into a profile of the dust-to-gas ratio. The absolute value of this profile however depends on the remnant dust accretion rate, which is initially higher due to radial drift, but more quickly decreasing than the gas accretion rate, as shown in Fig.~\ref{fig:dust_flux_t}.
\highlight{
The fact that the dust accretion rate is higher than the gas accretion rate multiplied by the initial dust-to-gas ratio is important for theories of planetesimal or planetary core formation because this allows a dust trap such as a vortex or zonal flow to achieve local dust over-densities. Depending on the trapping efficiency and the rate at which dust is accumulating in the trap, the dust over-densities can reach critical values to trigger the streaming instability and gravoturbulent formation of planetesimals, as discussed in \citet{Klahr:2006p7719} and \citet{Johansen:2006p7466}.
}

In the low turbulence case (top panels of Fig.~\ref{fig:dust_to_gas}) it can be seen that the disk regions inward of 4~AU get strongly enriched in dust because grains quickly grow to larger sizes and drift. This leads to a pile-up effect similar to \citet{Youdin:2004p15668}. This behavior can be understood by dividing Eq.~\ref{eq:sigd_frag} by the gas surface density, which results in a dust-to-gas ratio proportional to $r^{p-3/2}$. Since steady state $\alpha$ disks obey $p+q= 3/2$, we can also write
\begin{equation}
\epsilon_{frag} \propto T.
\label{eq:epsilon}
\end{equation}
Thus the steep increase in temperature due to viscous heating causes the increase in the dust-to-gas ratio through the temperature dependence of the gas disk profile.

In the high turbulent case (bottom panels of Fig.~\ref{fig:dust_to_gas}), the dust enhancement is much less effective because of the more effective fragmentation of dust grains due to higher turbulent velocities. This causes the drift velocities to be lower than in the low-turbulent case and for a given accretion rate, this yields only a modest enhancement of the dust-to-gas ratio.

It is important to note that so far we have considered the vertically integrated dust-to-gas ratio, which is what future observations may be able to constrain. For theoretical works, the mid-plane dust-to-gas ratio is usually more important. Due to settling, this will be even larger than the integrated value, if grains are large enough to efficiently settle down to the mid-plane. For particles with Stokes numbers $\alphat < \St < 1$, Eq.~\ref{eq:h_dust} describes the dust scale height as function of Stokes number reasonably well. For the low turbulence case, the particles at the drift limit (Eq.~\ref{eq:St_drift}) reach Stokes number of up to 0.1 at early times, later dropping $\lesssim$~0.01 (see also Fig.~\ref{fig:mass_distributions}). According to Eq.~\ref{eq:h_dust}, this means that the dust scale height and thus the mid-plane dust-to-gas ratio is enhanced by a factor of 3 to 10 compared to the vertically integrated values shown in Fig.~\ref{fig:dust_to_gas}. We find that at early times -- up to $3 \times 10^{5}$~years -- the mid-plane dust-to-gas ratio in the whole disk is increased at least by a factor of 2 over the canonical values. After that time, the loss of dust mass counteracts the settling effects. This result obviously depends on the initial condition of the dust-to-gas ratio, which in our case was taken to be a constant value of 0.01.

\section{Summary and conclusions}\label{sec:summary}
In this paper, we have derived a simple model that describes the radial evolution of the dust surface density under the combined influence of growth and fragmentation of compact grains as well as radial transport mechanisms. This has been achieved by finding upper limits to the grain size distribution, which are functions of time and the physical conditions at each radius. Good agreement between this very simple numerical model and a full-fledged, computationally intensive dust evolution code was found.

Important parameters are the fragmentation threshold velocity \uf, the level of turbulence \alphat, the initial dust-to-gas ratio, and the temperature and density profile of the gas disk. The effective dust transport velocity can be estimated by representing the dust distribution by only two characteristic grain sizes, the \emph{small} and the \emph{large population}. This allows us to derive analytical solutions for the dust surface density profiles in protoplanetary disks.

Our findings are summarized in the following:
\begin{itemize}
  \item The simple two-population model presented in this work describes the evolution of the dust-surface density and the evolution of the largest grain sizes well. Good agreement between this simplified model and a full-fledged dust evolution code was found. Additionally, the dust-to-gas ratio, the dust flux, and the size of the drifting particles can be derived from it.
  \item The upper end of the grain size distribution can be described as limited by turbulent fragmentation (cf., Eq.~\ref{eq:a_frag}) or by radial drift (cf., Eq.~\ref{eq:a_drift}). Fragmentation due to relative radial drift alone (Eq.~\ref{eq:St_df}) is found to be ineffective in non-laminar disks (see Section~\ref{sec:discussion:drift_velocities}). This supports the theory that disks are turbulent because radial drift alone cannot cause efficient fragmentation over long enough timescales to agree with observations. 
  \item We derived three different analytical profiles of the dust-surface density for different physical conditions: firstly, for a strongly drifting dust distribution (Eq.~\ref{eq:sigd_drift}), secondly for a fragmentation limited distribution (\ref{eq:sigd_frag}), and thirdly for a distribution where all grains are so small that they are well coupled to the gas (Eq.~\ref{eq:sigd_mixed}). The free parameter of the profiles is the dust accretion rate provided by the outer regions. The analytical profiles were found to fit to the simulation results of the full dust evolution code of \citet{Birnstiel:2010p9709} very well.
  \item We found that at late times of their evolution, disks can become drift-dominated, which for typical gas disk profiles ($\Siggas\propto r^{-1}$, $T\propto r^{-0.5}$) leads to a dust surface density profile proportional to $r^{-0.75}$. These results agree with the best-fit models for dust and gas profiles of the $\sim$10~Myr old TW Hya disk, as recently found by \citet{Andrews:2012p16676}.
  \item In the case of a fragmentation limited distribution with relatively large grains present, the dust surface density profile becomes proportional to $r^{-1.5}$ as in the Minimum Mass Solar Nebula \citep[see][]{Weidenschilling:1977p15694,Hayashi:1981p15696}.
  \item Similar to \citet{Youdin:2004p15668}, the vertically integrated dust-to-gas ratio is strongly enhanced in the innermost regions if the dust is significantly drifting in the outer region of the disk. This pile-up is aided by the combined effects of drift and fragmentation.
  \item An enhancement of the mid-plane dust-to-gas ratio was found in the low-turbulent simulation ($\alphat=10^{-3}$). This is because particles can reach large enough Stokes numbers (up to $0.1$) to efficiently settle to the mid-plane. In the case of efficient fragmentation, particles remain too small to significantly enhance the mid-plane dust-to-gas ratio.
  \item \highlight{As an important parameter for models of planetesimal or planetary core formation the radial mass flux in solids has been determined as a time and space dependent function. Depending on the disk parameters, the dust accretion rates can lie anywhere from a factor of a few up to two orders of magnitude above the value expected from the gas accretion rate, scaled by the dust-to-gas ratio. These values allow a dust trap such as a vortex or zonal flow to achieve a locally critical over-density on the order of only a few tens of orbits and could trigger the streaming instability and gravoturbulent formation of planetesimals, as discussed in \citet{Klahr:2006p7719} and \citet{Johansen:2006p7466}.}
\end{itemize} 

\begin{acknowledgements} 
We like to thank Satoshi Okuzumi, Chris Ormel, J\"urgen Blum, Anna Hughes, Phil Armitage, Sean Andrews, Luca Ricci for interesting discussions. We also thank the anonymous referee for a thorough and insightful report, which helped to improve this paper.
\end{acknowledgements}

\bibliographystyle{aa}
\bibliography{/Users/til/Documents/Papers/bibliography}

\makeatletter
\if@referee
\processdelayedfloats
\pagestyle{plain}
\fi
\makeatother
\appendix
\section{Simplified dust transport equation}\label{app:simplified_equation}
Assuming the gas to be the dynamically dominant medium, the radial transport of each trace species can be described by an advection-diffusion equation,  
\begin{equation}
\frac{\del \Sigma_i}{\del t} + \frac{1}{r} \frac{\del}{\del r} \left[ r \left(\Sigma_i \, u_i - D_i \, \Siggas \, \frac{\del}{\del r}\left( \frac{\Sigma_i}{\Siggas} \right) \right) \right] = 0
\end{equation} 
where $\Sigma_i$ is the surface density of the tracer, \Siggas the gas surface density and $u_i$ and $D_i$ are the tracer velocity and diffusivity, respectively.

In our case, the two trace species considered are the small and the large representative grain sizes, as described in Section~\ref{sec:two_pop_model}. The fraction of the total mass for each species is defined in Eq.~\ref{eq:sig_total}. Thus, the evolution of the total dust surface density can be written as
\begin{eqnarray}
\frac{\del \Sigdust}{\del t} &=& \frac{\del \Sigma_0}{\del t} +\frac{\del \Sigma_1}{\del t} \nonumber\\
                             &=& -\frac{1}{r} \frac{\del}{\del r} \left[ r \left(  \Sigma_0 \, u_0 + \Sigma_1 \, u_1 \right) \right] \\
                             &\phantom{=}&  +\frac{1}{r} \frac{\del}{\del r} \left\{ r \left[ \Siggas \, D_0  \, \frac{\del}{\del r}\left( \frac{\Sigma_0}{\Siggas}\right) +   \Siggas \,D_1 \, \frac{\del}{\del r} \left( \frac{\Sigma_1}{\Siggas} \right) \right] \right\}. \nonumber
\end{eqnarray} 

Using Eq.~\ref{eq:sig_total}, we can simplify this to an advection-diffusion equation for the total dust surface density
\begin{equation}
\frac{\del \Sigdust}{\del t} + \frac{1}{r} \frac{\del}{\del r} \left\{ r \left[\Sigdust \, u^* - D^* \, \Siggas \, \frac{\del}{\del r}\left( \frac{\Sigdust}{\Siggas} \right) \right] \right\} = 0,
\end{equation}
where
\begin{equation}
u^* = \bar u - \left( D_1-D_2\right) \cdot \frac{\del f_m(r)}{\del r},
\end{equation}
\begin{equation}
D^* = \left( D_1-D_2\right) \cdot f_m(r) + D_2,
\end{equation}
and $\bar u$ is the mass weighted velocity given by Eq.~\ref{eq:u_bar}. The dust diffusivities are \citep[see][]{Youdin:2007p2021}
\begin{equation}
D_i = \frac{D_\mathrm{gas}}{1+\St_i^2}.
\end{equation}
For Stokes numbers smaller than one, which is always fulfilled in this paper, the diffusivities are basically equal to the gas diffusivity and thus $D^* \simeq D_1 \simeq D_2$ and $u^* \simeq \bar u$. This allows us to simulate the radial evolution of the total dust surface density by using just the gas diffusivity and the mass weighted velocity, as in Eq.~\ref{eq:dust_simple}.

\section{Algorithm of the two-population model}\label{app:algorithm}
In this section, we summarize the algorithm of the two-population model.
\begin{enumerate}
  \item calculate the representative size for a fragmentation limited size distribution $a_\mathrm{frag}$ given by Eq.~\ref{eq:a_frag}.
  \item calculate the representative size for a drift limited size distribution $a_\mathrm{drift}$ given by Eq.~\ref{eq:a_drift}.
  \item calculate the representative size in the case of drift induced fragmentation
  \begin{equation}
  a_\mathrm{df} = \St_\mathrm{df} \, \frac{2\,\Siggas}{\pi\,\rhos}, 
  \label{eq:a_df}
  \end{equation}
  where $\St_\mathrm{df}$ given by Eq.~\ref{eq:St_df}.
  \item the representative size of the large population is now given by the smallest of the size limits
  \begin{equation}
  a_1 = \min(a_\mathrm{drift},a_\mathrm{frag},a_\mathrm{df})
  \end{equation}
  \item the initial phase where particles grow from the smallest size $a_0$ to $a_1$ is approximated by
  \begin{equation}
  a_1(t) = \min\left[a_1,a_0\cdot \exp\left(\frac{t-t_0}{\tau_\mathrm{grow}}\right)\right].
  \end{equation}
  where
  \begin{equation}
  \tau_\mathrm{grow} = \frac{\Siggas}{\Sigdust \, \Ok}.
  \end{equation}
  \item the velocities $u_0$ and $u_1$ of the two populations are then given by
  \begin{equation}
  u_{0/1} = \frac{\ugas}{1 + \St_{0/1}^2} +  \frac{2}{\St_{0/1}+\St_{0/1}^{-1}} \, u_\mathrm{drift},
  \end{equation}
  where
  \begin{equation}
  u_\mathrm{drift} = \frac{\csound^2}{2 \, V_\mathrm{k}} \cdot \frac{\mathrm{d}\ln P}{\mathrm{d}\ln r}
  \end{equation}
  is the drift velocity, $P$ the gas pressure at the mid-plane
  \begin{equation}
  P = \rhogas \, \csound^2 \approx \frac{\Siggas \, \Ok \, \csound}{\sqrt{2\,\pi}}
  \end{equation}
  and \St the Stokes number is defined by
  \begin{equation}
  \St_{0/1} = \frac{a_{0/1} \, \rhos}{\Siggas} \, \frac{\pi}{2}.
  \end{equation} 
  \item the effective dust transport velocity is the mass averaged velocity of both populations
  \begin{equation}
  \bar u = (1-f_\mathrm{m}) \cdot u_0 + f_\mathrm{m} \cdot u_1,
  \end{equation}
  where the mass fraction $f_\mathrm{m}$ depends on the size limiting mechanism
  \begin{equation}
  f_\mathrm{m} = \left\{\begin{array}{ll}
  0.97  &\text{if }a_\text{drift}\text{ is the smallest of }a_\mathrm{frag}, a_\mathrm{drift}, a_\mathrm{df}\\
  0.75  &\mathrm{otherwise}\\
  \end{array}
  \right.
  \end{equation}
  \item Having calculated the effective transport velocity $\bar u$, the evolution of the dust surface density can be simulated by numerically solving the advection-diffusion equation, Eq.~\ref{eq:dust_simple}.
\end{enumerate}
\end{document}